# The Rise of Ingot Niobium as a Material for Superconducting Radiofrequency Accelerating Cavities

P. Kneisel,[1] G. Ciovati,[1] P. Dhakal,[1] K. Saito,[2] W. Singer,[3] X. Singer,[3] and G. R. Myneni[1]

[1]*Jefferson Lab, Newport News, VA 23606, USA*
[2]*Michigan State University, East Lansing, MI 48824, USA*
[3]*DESY, Notkestrasse 85, D-22607 Hamburg, Germany*

*Abstract*

As a result of a collaboration between Jefferson Lab and niobium manufacturer CBMM, ingot niobium was explored as a possible material for superconducting radiofrequency (SRF) cavity fabrication. The first single cell cavity from large grain high purity niobium was fabricated and successfully tested at Jefferson Lab in 2004. This pioneering work triggered research activities in other SRF laboratories around the world. Large grain niobium became not only an interesting alternative material for cavity builders, but also material scientists and surface scientists were eager to participate in the development of this material. Most of the original expectations for this material of being less costly and allowing less expensive fabrication and treatment procedures at the same performance levels in cavities have been met. Many single cell cavities made from material of different suppliers have been tested successfully and several multi-cell cavities have shown the performances comparable to the best cavities made from standard poly-crystalline niobium. Several 9-cell cavities fabricated by Research Instruments and tested at DESY exceeded the best performing fine grain cavities with a record accelerating gradient of $E_{acc}$ = 45.6 MV/m. Recently- at JLab- by using a new furnace treatment procedure a single cell cavity made of ingot niobium performed at a remarkably high $Q_0$-value (~$5\times10^{10}$) at an accelerating gradient of ~20 MV/m, at 2K. Such performance levels push the state-of-the art of SRF technology to new limits and are of great interest for future accelerators. This contribution reviews the development of ingot niobium technology and attempts to make a case for this material being the choice for future accelerators.

## 1. INTRODUCTION

Niobium was introduced as material for superconducting accelerating cavities in 1967, replacing the lead-plated copper cavities, which had been used at High Energy Physics Lab (HEPL), Stanford University, Kernforschungszentrum Karlsruhe (KfK), Brookhaven National Lab (BNL) and CERN. Initially, the electrolytic deposition of niobium from a salt bath onto a copper substrate was used [1]; however, because of the high deposition temperature of 740 $^{o}$C the layers were contaminated by diffused Cu ions. Later, Siemens AG succeeded to fabricate a $TE_{011}$ X-band cavity by this method, but with low $Q_0$ values [2]. At Stanford University, which was the pioneering laboratory in the development of niobium cavities, the first results were achieved in 1968. Starting with vapor deposition of thin layers and electro-deposition the research quickly turned to using solid reactor grade niobium and machining X-band cavities from ingot material [3,4,5]. This route was also taken at Siemens AG [2] and at KfK for R&D cavities and separator modules [7]. Figure 1 shows the several examples of cavities made from ingot material/electro-deposition.

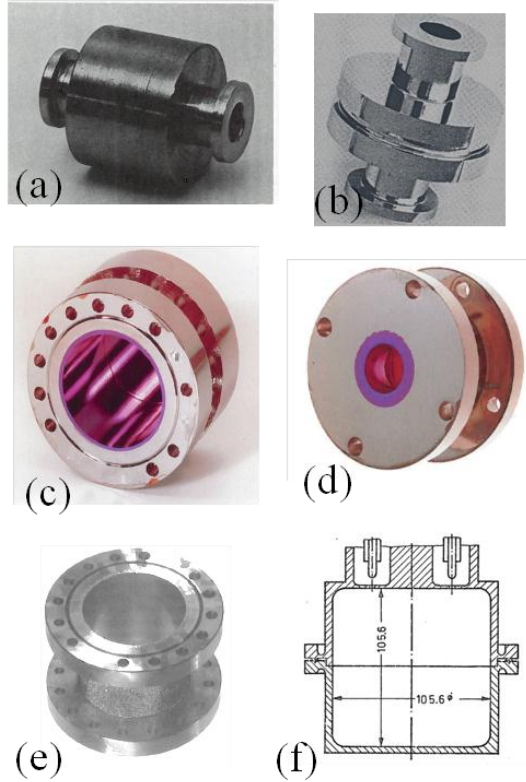

Figure 1: Niobium Cavities from Ingot Niobium: (a) and (b) are manufactured at HEPL, $TM_{010}$ cavity [4, 6], (c) $TE_{011}$ cavity [2], (d) $TM_{010}$ cavity at Siemens and anodized, cavity (e) manufactured by electro-deposition at Siemens and (f) is the schematic of a $TM_{011}$ cavity manufactured at KfK [7]

These cavities from ingot, reactor grade material (a)-(d) performed well with peak surface magnetic fields as high as 107 mT [6] and 159 mT [8]. Following the excellent results obtained with X-band cavities at HEPL, multi-cell L-band cavities for the Superconducting Linear Accelerator were designed and fabricated. Results from several cavities made from hydro-formed sheet metal discs were reported in ref. [9]. The manufacturing technology developed at HEPL in the 1970s used the deep drawing of half cells, machining to final dimensions and electron beam welding at the equators and irises; it was copied by every cavity fabricator since then with only minor modifications such as elimination of the final machining step, and niobium sheet metal became the material of choice. Since then several accelerator projects rely on the SRF cavities and fabrication technology based on the high purity Nb sheet or sputtered Nb on copper cavities.

The high purity Nb sheets which are used in the standard fabrication of SRF cavities have a grain size of ~50 μm (ASTM ≥ 5). A small and uniform grain size was considered to be a necessary requirement for good formability of polycrystalline material to avoid an excessive roughness of the surface after deep-drawing (having the appearance of an "orange peel"), resulting from the non-uniform deformation of large grains with different crystalline orientation. The production of fine-grain Nb sheets involves many steps, including forging, rolling and heat treatments, to break the large crystals of the Nb ingot, which had been melted multiple times, into uniform fine-grains.

The idea of using ingot material for cavity fabrication in the form of sheets sliced directly from an ingot and shaping the slices with the standard deep drawing technology, rather than carving out the half cells from thick chunks of material as was done for the separator cavities at KfK, was introduced at Jefferson Lab in 2004 with the US patent filed as "large grain cavities from pure niobium ingot" [10]. It turned out that this sliced material had very good mechanical properties and could be easily formed into half cells despite the large grains (the size of the grains is few cm). This review article includes the summary of several studies on material characterization (Sec. 2), cavity fabrication (Sec. 3) and testing (Sec. 4) around the world based on single crystal and large grain niobium. In Section 5 we will discuss the achievements and some open issues related to the ingot Nb technology and a brief summary will be given in Sec. 6. Additional information on the topics discussed in this article can be found in several conference proceedings on SRF technology [11] and niobium materials for more details [12, 13].

## 2. MATERIAL STUDIES

### 2.1 The Promise of Ingot Material

Once it was determined that ingot slices could be used for SRF cavity fabrications, it became obvious that this choice of material could have certain advantages over standard poly-crystalline material due to:
- Reduced cost because of elimination of sheet processing and less waste.
- Comparable or better performance.
- Possibility of higher Q-values/lower residual resistance because of less grain boundaries, reducing the cryogenic load and therefore lowering operational costs of SRF accelerators.

- Smooth surfaces with buffered chemical polishing (BCP) instead of electropolishing (EP) because of less grain boundary etching.
- Higher thermal stability because of a phonon peak in thermal conductivity.
- Less material quality assurance efforts (eddy current scanning) because of elimination of rolling and intermediate heat treatments, which can introduce defects
- Better reproducibility of performance because of lower defect density.
- Good or better mechanical reproducibility of predictable spring back after forming.

However, there are some issues connected with this material:
- Can one develop an effective/inexpensive slicing method?
- How uniform is the forming process, is slippage at grain boundaries problematic?
- Do grain boundaries cause a problem during welding?
- Are grain boundaries leak tight?
- Are mechanical properties depending on crystal orientation?
- Are different grain orientations reacting differently to chemical etching?
- Is oxidation behavior depending on grain orientation?
- Is field emission grain boundary dependent?

Many of the issues listed above are technological and provided a rich field of research for material scientists and surface scientists. Table 1 summarizes the different areas of research which have been addressed in the last 8 years. Some of the results from this research are discussed in more details in the following sections.

Table 1: The Summary of the material studies conducted on ingot Nb in recent years.

| Subject | Comments | References |
|---|---|---|
| Mechanical Properties | Yield strength, tensile strength, elongation for different grain orientation, bulging, residual strain, formability | [14,15,16,17,18,19] |
| Thermal properties | Thermal conductivity, phonon peak, effect of annealing and impurities, effect of strain | [14, 20,21,22] |
| Magnetic/electrical properties | $H_{c1}$, $H_{c2}$, $H_{c3}$ for different crystal orientations, temperature dependences, penetration depth | [21,23, 24, 25] |
| Crystal orientation, recrystallization | Grain orientation in different materials, dislocation density, dependence of etch rate on orientation and residual strain | [26,27] |
| Flux penetration | Magneto-optical investigations, influence of grain boundaries on flux penetration | [24, 25, 28] |
| Oxidation / Hydrides | Oxide composition for different crystal orientations, sealing of surface, hydrides | [29] |
| Field emission | Emitter density, grain boundary segregation, cleaning | [30, 31] |
| Fabrication | Forming issues, EBW, enlargement of single crystal, avoid recrystallization, recovery | [17, 32] |
| Surface topography | Influence of polishing conditions, replica | [33, 34, 35, 36] |

|  | technique, pits, roughness, field enhancement |  |
| --- | --- | --- |
| Hot spots, cold spots | Point contact tunneling, vacancies, dislocations | [37,38] |
| SIMS analysis | Hydrogen depth profile | [39, 40] |

### 2.2 Mechanical and Thermal Properties

The large grain material originally provided by Companhia Brasileira de Metalurgia e Mineração (CBMM), Brazil, as part of the cooperative research and development agreement had excellent mechanical properties as shown in Fig.2 for a single crystal piece cut from a large grain ingot compared to the polycrystalline niobium. Figure 3 shows the stress-strain curves for different samples with different crystal orientations from a sheet of ingot material. Despite of grain boundaries, the elongation is quite high.

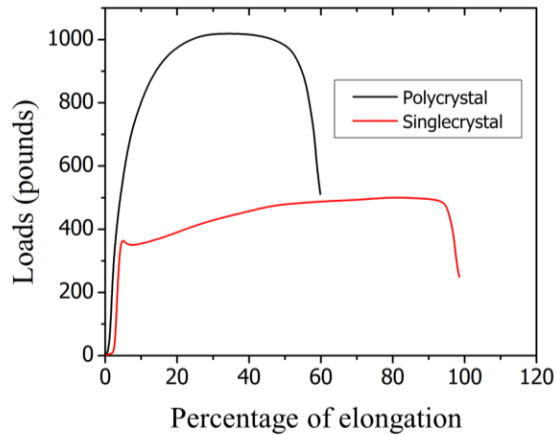

Figure 2: Stress-strain behavior of poly-crystalline and single crystal niobium [taken from ref. 32].

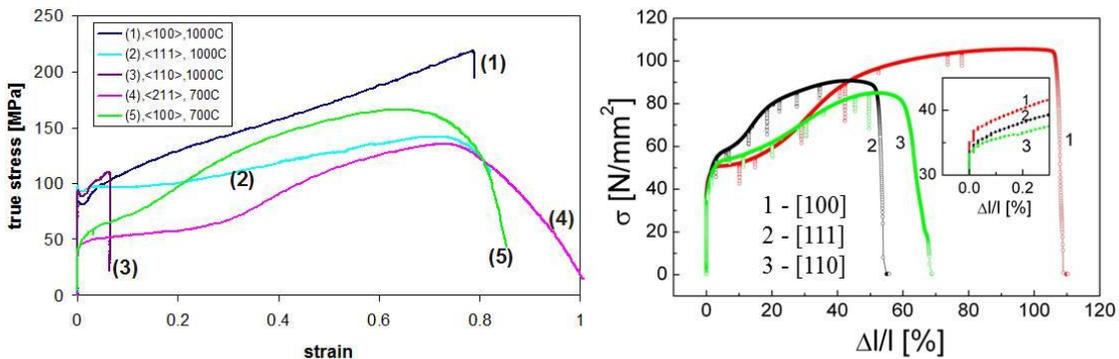

Figure 3: Left: stress-strain curves for three different samples with different crystal orientations [100], [111], and [110] from a sheet of ingot material [16]. Despite of grain boundaries, the elongation is quite high [17].

As part of the development program at CBMM for high purity niobium ingots, several ingots have been produced with different parameters during the melting process. The mechanical properties of samples cut from ingots with different Ta content were measured after different numbers of melting cycles [41]. The results showed that the ingots with high and low tantalum content which had greater than four melts showed lower mechanical strength and greater elongation compared to ingots after third melt cycle.

More stringent testing of the material properties relevant to forming into cavity cells and conforming to the pressure vessel code is done with a bi-axial bulging test. Such testing was done at DESY (Fig. 4): the percentage elongation to fracture is low (<15%) and the rupture takes place close to a grain boundary. Although this might have some impact on the formability of large grain niobium sheets, no major difficulties have been encountered in the fabrication of cavities from this material, as described below.

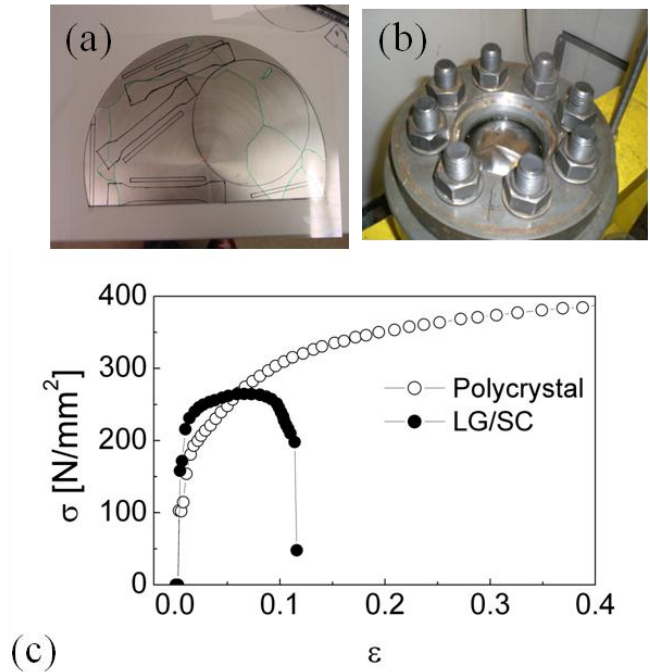

Figure 4: Bi-axial bulging test. (a) The disc with 3 grains of [100] and [111] orientation used for the test, (b) shows the ruptured disc in its holder and (c) the result of the test [17].

Mechanical properties of large grain/single crystal ingot material have been further investigated at MSU. Microstructure, dislocation density, texture and slip behavior has been studied extensively. Based on the measured texture of large grain samples, the forming processes can be computationally simulated for these grain orientations [18].

The thermal conductivity of both single crystal and large grain niobium was measured at several labs and Fig. 5 shows results obtained at JLab [14], DESY [17] and MSU [20].

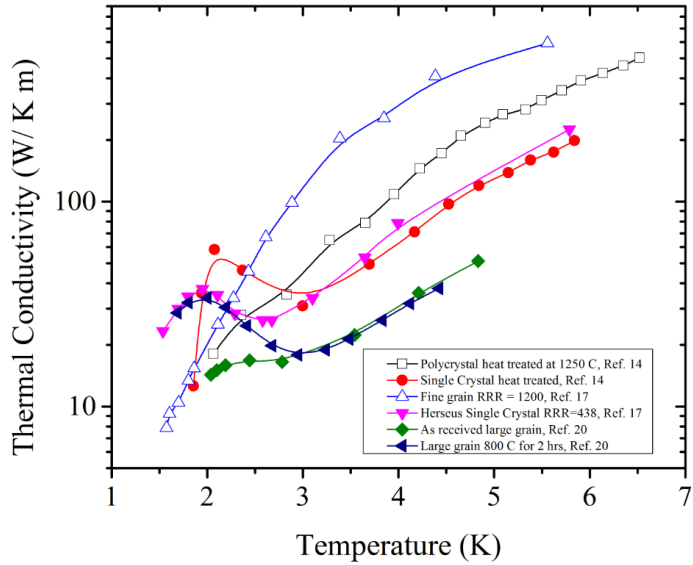

Figure 5: Thermal conductivity of single crystal, large grain ingot niobium compared to fine grain material. There is usually a phonon peak detectable in ingot niobium.

At MSU the thermal behavior of large grain material was further investigated with respect to the dependence of the phonon peak on mechanical deformation, uptake of hydrogen and different RRR values. As it has been reported in ref. [22], the phonon peak is very sensitive to mechanical crystal strain, but can be recovered by annealing heat treatments. In the range between 800 °C and 1200 °C the dislocation density was reduced by approximately a factor of 4 and accordingly the phonon peak heights increased about 2.5 times, exceeding the value of thermal conductivity at 4.2 K. Hydrogen in niobium also affects the thermal conductivity especially in the phonon peak region, but a full recovery can be achieved after degassing at 800 °C for 3 hrs. The phonon peak height also increases with increasing RRR value. These results are shown in Fig. 6.

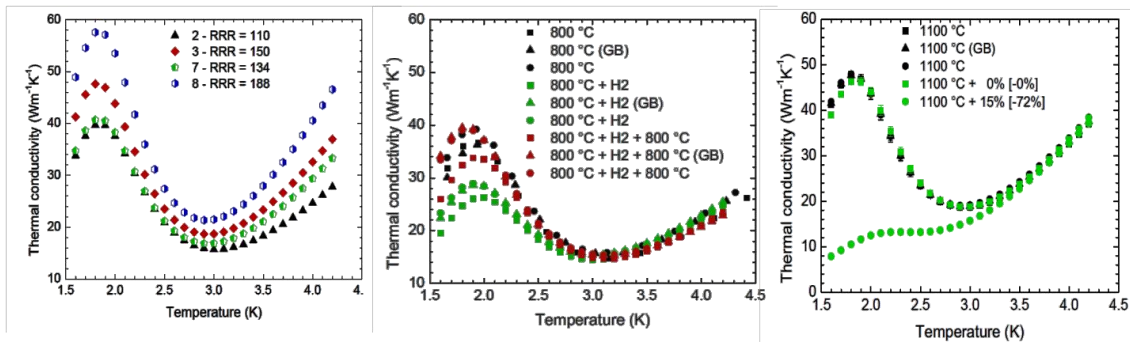

Figure 6: Dependence of the phonon peak on RRR-value (left), hydrogen concentration and recovery by heat treatment at 800 °C (middle) and on strain in the material and recovery by heat treatment (right) [22].

## 2.3 Formability, Crystal Deformation and Crystal Structure

For material scientists it is of great interest to understand how the crystal structure is responding to severe deformations such as those applied during deep drawing of cavity half-cells. During the deep drawing process a plastic strain differential is created through the material, which induces residual stresses. Their magnitude depends on yield stress, the accumulated plastic strain and the hardening of the material, all influencing the formability of the material and spring back behavior after forming. First measurements of residual stresses and formability of large grain and single crystal niobium were reported in [16]. Because of the large strain reserve as shown in Figs. 2 and 3 for uni-axial testing and an estimated maximum local strain of ~ 20% there is a large margin even for the biaxial nature of the forming. However, anisotropy in yielding provides a potential for local thinning and shape deviations. This has been observed at many places, especially slippages at grain boundaries.

## 2.4 Impurities in Surface Layer of Ingot Niobium, Oxides and Hydrides

Sheets from ingot material usually consist of several large grains with decreasing grain size towards the outer circumference (this depends on the size of the water-cooled crucible during the melting process). These grains have typically different crystal orientations and react differently to mechanical deformations, to chemical agents such as buffered chemical polishing solutions and show also different oxidation behaviors. Several investigations as discussed below have addressed these issues. Single crystal niobium samples of [100], [110], and [111] crystal orientations were analyzed using TEM and SIMS techniques [29] For all 3 orientations the uniform oxide layer thickness ranges from 4.9 to 8.3 nm (only one sample of each orientation was analyzed) and no significant sub-oxides were seen. Figure 8 shows cross-sectional TEM micrographs of the oxide layer on the 3 different orientations.

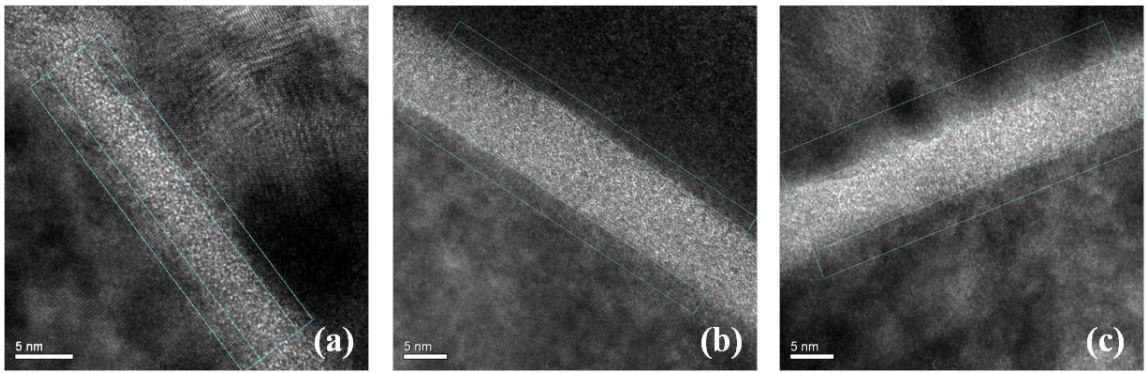

Figure 7: Cross sectional TEM micrographs of (a) [100], (b) [110], and (c) [111] single crystal niobium showing the niobium oxide layer [29].

At the Surface Science Lab of Krakow University the oxidation behavior of single crystals of different orientations were studied with XPS and compared to the oxides on poly-crystalline niobium after buffered chemical surface treatment [42]. The XPS spectra revealed that the oxide layer on fine grain niobium is thicker than on single crystal material.

SIMS measurements were made on samples of all 3 orientations to obtain depth profiles of hydrogen, carbon and oxygen after heat treatments at 90 °C, 600 °C and 1250 °C. High levels of hydrogen were found between the oxide layer and the niobium, but low levels in the oxide. Below the oxide layer no high oxygen concentration regions were found and carbon contamination was mainly detected on the surface. The analysis of interstitial elements in the surface of niobium with ion mass spectroscopy continued at North Carolina State University with particular interest in the concentration profiles of hydrogen. Figure 8 shows a dramatic decrease of the hydrogen concentration after a sample was subjected to a heat treatment at 800 °C for 3 hrs followed by a 120 °C bake for 24 hrs [39].

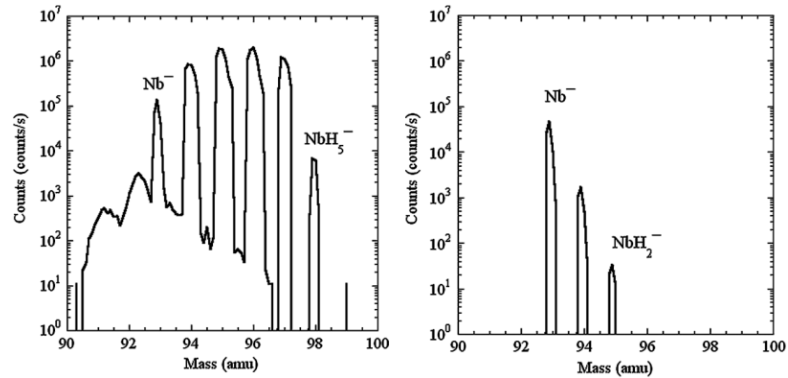

Figure 8: SIMS mass spectra of the hydrogen concentration in a (a) non-heat treated and (b) heat treated single crystal niobium sample. The heat treatment was done at 800 °C for 3 hrs and subsequent "in situ" baking at 120 °C for 24 hrs [39].

The picture which emerged from these investigations is as follows: during a heat treatment at sufficiently high temperature the oxide surface layer is dissolved into the niobium matrix and allows desorption of the hydrogen from the bulk. With exposure to air a dense oxide layer is forming, which then prevents the absorption of hydrogen from the atmosphere. This method was already applied successfully as reported in ref. [43] to prevent hydrogen pick-up after UHV degassing of niobium. A comparison between high RRR fine grain and large grain/single crystal surface composition showed that with the same treatments no differences were found for C, O, and N but that hydrogen was present in higher concentrations in poly-crystalline niobium after the degassing. These observations may point to hydrogen as a major impurity influencing rf performance of cavities with different crystallographic structure.

The detrimental effect of large hydrogen concentrations in high purity niobium on rf cavity performance were experienced long time ago in the late 1980's [44] and have become known as "Q-disease" with the signature of strong Q-degradations at very low fields. The physical effect taking place is the precipitation of dissolved hydrogen introduced into the material during manufacturing and surface treatments into the NbH-ε–phase at a temperature around 100 K; this phase is weakly superconducting and therefore is enhancing the losses in the material. Cures to this problem have been either a fast cool down of cavities through the dangerous temperature region of 75 K < T < 150 K to avoid the formation of the NbH phase or to hydrogen degas the cavities at temperatures > 600 °C in vacuum for several hours. Since the rf losses in a cavity take place only in the penetration depth (~ 50 nm), the hydrogen had to be concentrated in this surface sheath- this has been verified with different surface analytical tools such as SIMS, Nuclear Micro Probe, ERDA and concentrations of 10 to 50 at.% have been detected, orders of magnitude higher than in the bulk. Hydrogen is very mobile in niobium and tends to accumulate near defects – the surface is a major defect. Hydrogen trapping also occurs near interstitial or substitutional impurities, retarding hydrogen precipitation in the ε–phase. This might be the cause for the absence of Q–disease in low RRR niobium cavities.

After the discovery of the "Q–disease", hydrogen has become again the focus of attention: this time in the context of finding the causes of the so-called "Q–slope" and "Q–drop" in high performing niobium cavities. In short, the $Q_0$ vs $E_{acc}$ cavity performances show a slight degradation of Q–value in the gradient range of $5 < E_{acc} < 25$ MV/m range and a dramatic drop in Q–value above this field. Additionally, these non-linearities in $Q_0$ vs $E_{acc}$ can be eliminated to a large extend by baking the cavities at a moderate temperature of ~ 120 °C for > 12 hrs (LTB). In ref. [45], some of the earlier experiments with ERDA were repeated employing the presently used surface preparation techniques for cavities. Samples were high purity single crystals, large grain and poly-crystalline pieces. The samples were – after different treatments – studied in ERDA setup using a 1.6 MeV $^4$He beam. Data were taken at different spots on a sample and the variation from spot to spot was below 5 % indicates a uniform distribution of hydrogen. Even though in all samples after different treatments hydrogen peaks were detected in a near surface sheath of ~ 8 nm, of the order of the depth resolution, no difference in hydrogen depth profile was seen after LTB or between samples cut from cavities at locations of high and low rf losses (referred to as "hotspots" or "coldspots"). Because ERDA does not distinguish between interstitial hydrogen and precipitated hydrogen, a possible explanation has been given by assuming different states of precipitation on hot and cold areas in a cavity, initiated by very different concentrations of lattice defects and vacancies.

Recent investigations of surface hydrogen and oxygen concentrations on niobium surfaces using Raman spectroscopy [46] identified ordered hydride phases near large concentrations of crystal defects such as dislocations and vacancies, which had shown up as "hot" spots in cavities and as etch pits in the surface topography. These etch pits will likely attract sufficient hydrogen to form lossy hydrides with suppressed superconductivity.

### 2.5 Chemical Etching and Surface Finish

Superconducting niobium cavities typically undergo a surface treatment to remove the "surface damage" layer after fabrication. It has been shown in series of tests with poly-crystalline niobium treated by BCP that the "damage layer" is of the order of 100–150 μm thick [47]. The crystallographic structure however is already restored after ~10 μm material removal. Other methods used in the preparation of rf cavities are EP and/or CBP and for very high accelerating gradients these methods have been superior to BCP.

Since one of the expectations for large grain material is less expensive preparation procedures, some emphasis has been placed on the achievable surface finishes with different methods. It has been shown for poly-crystalline niobium that BCP treatment causes surface roughness due to grain boundary etching, which in some cases has been blamed for inferior rf performance of cavities. Large grain or single crystal niobium exhibits also preferential, but less pronounced etching at grain boundaries, however mirror like surface finishes can be achieved on individual grains exhibiting very smooth surfaces as shown in Fig. 9.

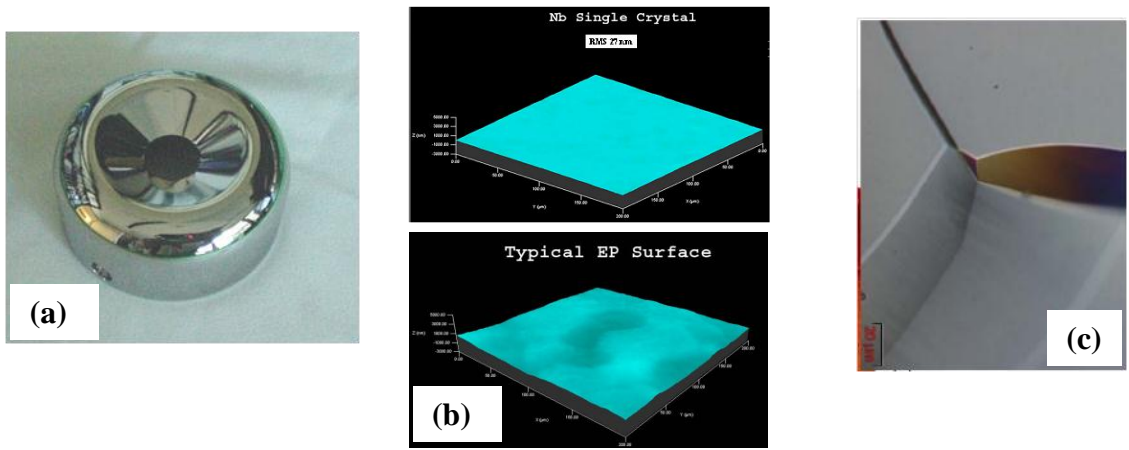

Figure 9: Examples of single crystal surfaces after BCP of ~ 100 micron material removal; shown is a (a) single crystal cathode for an electron injector, (b) a surface profile from a flat single crystal sample with an average roughness of 27 nm of the samples area compared to an EP surface [48], and (c) an etched surface of a large grain triple junction.

Why seems the issue of surface finish to be so important?

For long time it has been promoted that very smooth cavity surfaces are essential to achieve excellent performances since the super-current flows in a thin surface layer of app. 50 nm and topographic features of the same scale would matter. On this subject, it should be mentioned that there are many examples of non-ideal surfaces which look rough and damaged to the naked eye, but performed excellently [49]. Nevertheless, there is quite some effort invested in investigating surface topography under varying alternative treatment conditions and for different materials. In refs. [50, 51] topographical changes of niobium under controlled EP conditions are reported. In these investigations, samples of high purity poly-crystalline, large grain and single crystal niobium were subjected to comparative BCP, centrifugal barrel polishing (CBP), and EP treatments. The surfaces were examined with stylus profilometry (SP), optical microscopy and Atomic Force Microscopy (AFM) and the data from several scanned locations of the samples were combined into a Power Spectral Density (PSD) plot of the surface roughness after a Fourier transformation of the measured surface heights had been carried out ( see details in ref [33]). The study concludes that nano-scale surface roughness on small areas can best be achieved by applying an optimized CBP process to "standardize" a surface followed by a moderate EP of > 30 μm.

The response of poly-crystalline and single crystal/large grain niobium to BCP was further investigated in ref. [52] by examining the topographical features by optical microscopy after the polishing procedure. Of particular interest in the context of this review is the observation that nano-polished samples initially showed smoothening, but also deepening of the grain boundaries. But further etching created dents in the surface as can be seen in Fig. 10.

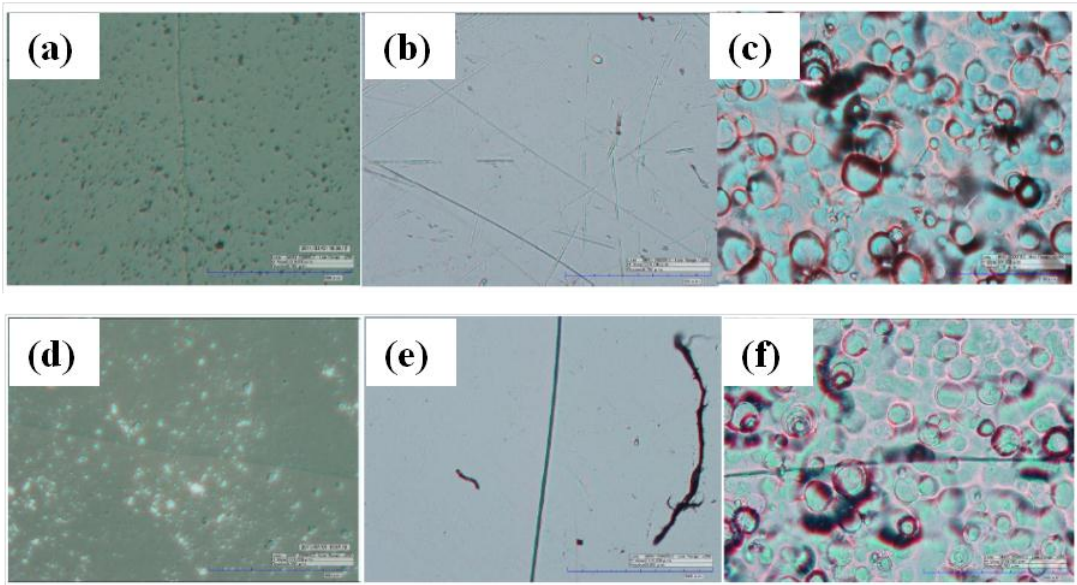

Figure 10: Single crystal: (a) nano-polished, (b) BCP of ~ 30 μm exposing scratches from the mechanical polishing and (c) ~ 80 μm BCP. Bi-crystal: (d) nano-polished, (e) BCP of ~ 30 μm and (f) ~ 80 μm BCP. Preferential etching at the grain boundary can be seen in (e) [52].

### 2.6 Grain Boundary Flux Penetration

Superconducting cavities are limited in their high field behavior in two ways: The superconducting material makes a thermal transition from the superconducting state to the normal conducting state ("global heating") or the material reaches its local critical field and "quenches". In that case magnetic flux has entered the material and has driven the material into the normal conducting state. This latter transition is in nearly all cavities the limiting factor.

Some of the weak areas in a cavity surface are the grain boundaries, which are chains of dislocations, where flux can enter preferentially. Using magneto-optical imaging (MO) it was shown on fine grain niobium [28, 53] that some grain boundaries indeed showed preferential flux penetration at a field lower than that for which flux penetration occurs within the grains, the first experimental evidence of depressed superconductivity at a grain boundary in Nb. This study had left some questions open to what extent did surface topology and chemical surface treatments, as well as heat treatments to enlarge the grain size, influence the flux penetration behavior. Large grain/single crystal niobium provided the opportunity for "cleaner" conditions to investigate the mechanism of flux entry at a grain boundary and flux flow along a grain boundary. Such investigations have been carried out on bi-crystal and tri-crystal large grain niobium samples [54]: in a bi-crystal with < 1 μm step at the grain boundary and the grain boundary plane nearly parallel to the applied magnetic field preferential flux entry was observed at field levels far below $H_{c1}$, whereas at large steps and/or grain boundaries not parallel to the field no preferential penetration was observed, excluding topological features causing this behavior. A continuation of this work showed that flux penetration will also occur at the "stable" boundary, if it is rotated such that it is parallel to the applied field. What causes these weaknesses? TEM images suggest that dense dislocation arrays at the sample surface possibly introduced by the preparation techniques such as cutting and grinding, are responsible for the reduced $H_{c1}$. In the case of ingot material the density of grain boundaries is lower, but the length of these grain boundaries is also longer, therefore it is difficult to assess its advantage compared to fine-grain niobium, especially since the grain boundaries in formed cavity cells are randomly oriented with respect to the rf field.

    Measurement of the depairing current density is an additional method to characterize weak areas in grain boundaries. If grain boundaries locally depress the superconducting energy gap, they will also depress the depairing current density and enhance the flux flow resistivity. As described in detail in ref. [54] grain boundaries show preferential flux flow when the external magnetic field is almost parallel to the plane of the grain boundary. Surface treatments such as BCP or EP might reduce locally the energy gap at the grain boundary or reduce pinning of vortices along grain boundaries. The comparison of both BCP and EP treatments indicated that BCP has a more serious effect on deteriorating the superconducting properties at grain boundaries.

## 2.7 Magnetization, Critical Fields, Penetration Depth, and RF Properties

In the early 1970's a program was started at KFZ Karlsruhe to develop new methods to investigate "SRF grade" niobium using the conduction electrons as a probe to sample material properties at different depth in the material. Volume properties such as residual resistivity and bulk mean free path were investigated with dc and 10 kHz penetration depth measurements. Measurements of flux pinning and magnetization ($H_{c2}$) explore depths between 1 and 50 μm, whereas penetration depth measurements as a function of a magnetic field parallel to the niobium surface probe a depth of 0.04 to 1 μm and are useful to determine surface mean free path and sub-oxide precipitates. The niobium/oxide interface can be sampled with $H_{c3}$ and tunnelling measurements to a depth of ~40 nm [55]. This type of measurements have been recognized in recent years as a powerful method to gain information about surfaces subjected to different cavity treatments such as BCP, EP, heat treatment, and "in-situ" baking [21, 22, 25].

Magnetization measurements give information about bulk properties such as magnetic field for first flux penetration ($H_{ffp}$), $H_{c2}$ and irreversibility, related to pinning centers in the material The magnetization measurements were carried out on 120 mm long large grain cylindrical rods of different diameters as well as the different RRR and impurities contents. The results showed that surface and heat treatment have little influence on the first flux penetration field ($H_{ffp}$ ~ 170-190 mT), whereas the hysteresis is significantly reduced after the 800 °C heat treatment. Furthermore, the critical current density and pinning force calculated from magnetization data indicate that LG samples have lower critical current density and pinning force density compared to FG samples, therefore favoring lower flux trapping efficiency [25]. This effect may explain the lower values of residual resistance often observed in LG cavities than FG SRF cavities [56].

In ref. [57] detailed magnetization measurements on single crystal and large grain samples, 2 × 2 × 2 mm$^3$ in size, were reported to establish the critical field for first flux entry; this was done on identical samples of "pristine" condition (as received from the vendor), and on BCP and annealed samples. As shown in Fig. 11, first flux entry in the "pristine" sample is delayed up to ~200 mT; the samples display a large hysteresis loop and several flux jumps occur. After BCP and heat treatment, the hysteresis is significantly reduced, flux jumps are absent and first flux entry occurs around 90–100 mT. This result is consistent with a reduction of pinning centers in the surface after the BCP surface treatment as well as heat treatment and it has been argued that bulk pinning might also be effected by BCP treatment due to interactions between hydrogen absorbed during chemical treatment and vacancies, dislocations and Ta impurities in the niobium. The value of $H_{ffp}$ is very low compared to reported value of $H_{c1}$ in the literature for high-temperature annealed samples [58]. The causes for differences on the dependence of $H_{ffp}$ on surface treatments measured on the small, cube-shaped samples versus that obtained from cylindrical rod-shaped samples are not clear. This might be due to different instrument sensitivities or to the different surface-to-volume ratio between the two types of samples.

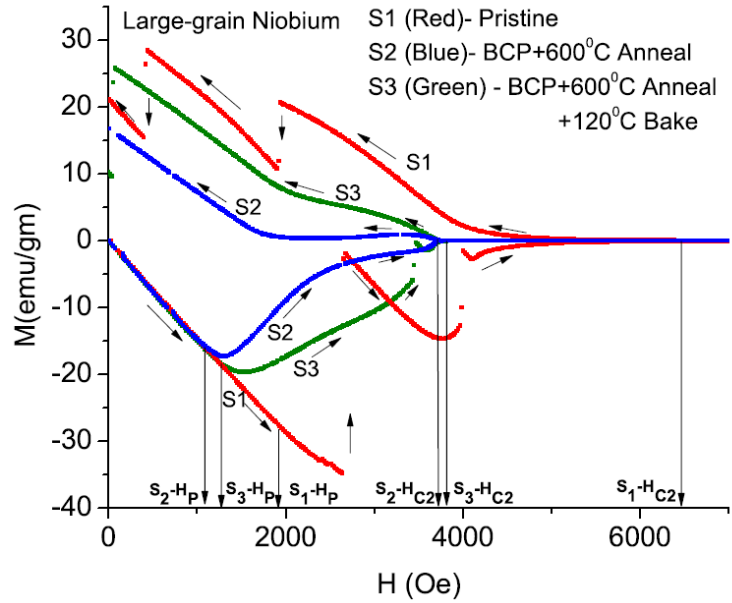

Figure 11: Isothermal magnetization at 2 K of large grain niobium "pristine" and BCP and heat treated samples [57].

Penetration depth measurements as a function of temperature give information about the mean free path in the probed layer (for 330 kHz it is ~ 25µm); as a function of external magnetic field it is a sensitive measurement of the magnetic properties in the surface layer up to $H_{c3}$. These investigations are carried out as the measurement of the frequency shift of a 300 kHz LC oscillator [59], in which the coil – the inductance of the circuit – was tightly "filled" with the sample under test. This setup is very sensitive to changes of material surface properties due to different treatments and surface superconductivity ($H_{c3}$) can still be detected, when the bulk of the sample is already normal conducting ($H_{c2}$). Figure 12 shows the change in resonant frequency, which is proportional to the change in penetration depth, of the LC oscillator as a function of external applied DC magnetic field at 2 K for large-grain and fine grain samples after ~30 µm EP and LTB [25]. The increase in $H_{c3}$ after LTB is consistent with earlier results [233]. Work is in progress to develop a model to describe $\Delta f(H)$ in the region between $H_{c2}$ and $H_{c3}$ as a function of material properties.

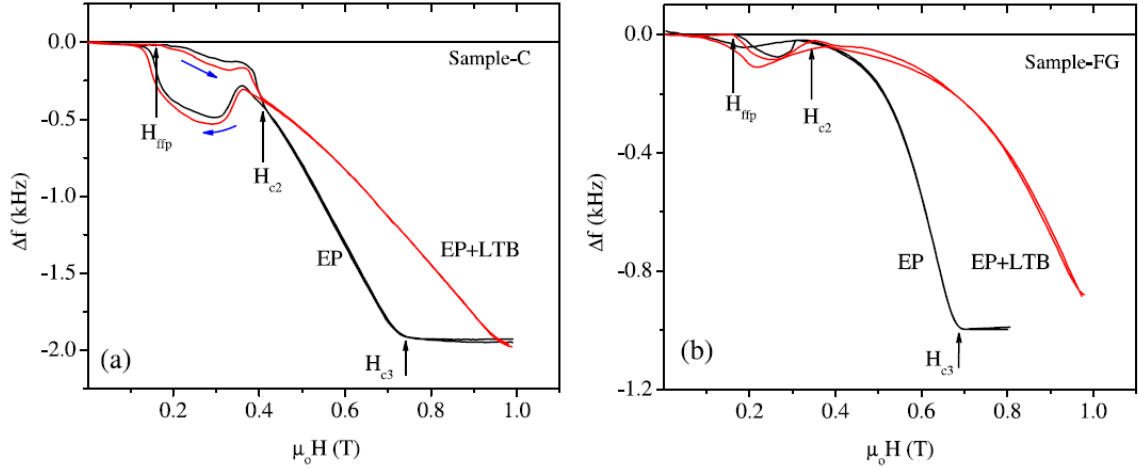

Figure 12: Change in resonant frequency at 2 K of the LC oscillator as a function of the external magnetic field showing the transition at $H_{ffp}$, $H_{c2}$ and $H_{c3}$ for a large-grain sample (a) and a fine-grain sample (b) subjected to ~30 μm EP and LTB [25].

In attempting to correlating dc and low-frequency measurements to the rf superconducting properties of Nb samples, a coaxial cavity operating in the $TE_{011}$ mode at 3.5 GHz was developed at Jefferson Lab [60]. The latest version of the cavity and the sample are shown in Fig. 13.

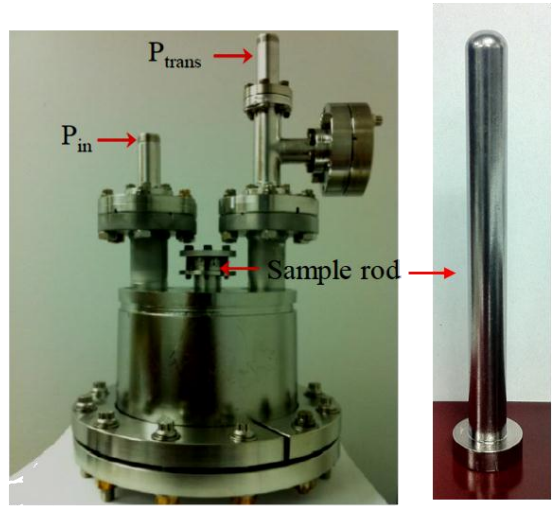

Figure 13: Coaxial cavity and test set-up for measuring the superconducting properties of the coaxial sample.

RF measurements of the samples are done by inserting them coaxially into the rf cavity shown in Fig. 13. The maximum magnetic field is in the middle of the rod and is 2.2 times higher than everywhere on the pillbox surface. Prior to the sample test, the cavity is measured empty to ensure that the quench field of the empty cavity is larger than the quench field in the coaxial configuration. In this case the empty cavity is limited at $B_{peak}$ = 78 mT with $Q_0 = 5.8 \times 10^9$ at 2 K. Figure 14 shows the rf measurement on one of the large grain samples after different surface treatments as indicated in [21]. The magnetic peak field on the sample was limited to $B_{peak}$ = 40 mT due to critical heat flux of liquid He at 2.0 K through the cooling channel in the sample. A sample from a different ingot subjected to EP surface treatment followed by heat treatment at 600 °C for 10 hrs and LTB at 120 °C for 12 hours has the higher Q-value with a quench field of $B_{peak}$ ~ 55 mT. Additional tests are planned to fully characterize the rf properties of the ingot samples and correlate them with the rf performance of cavities.

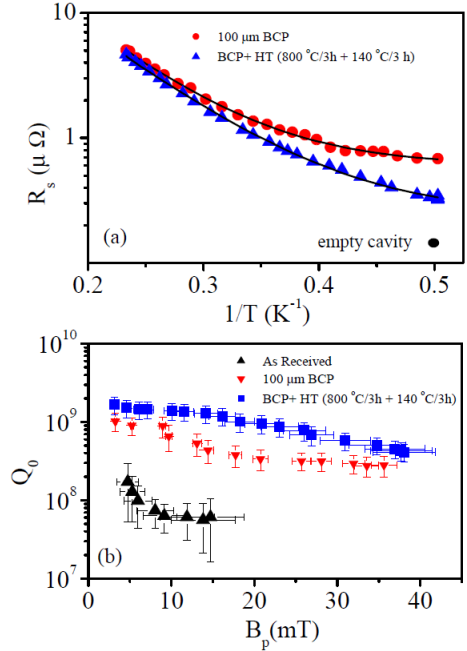

Figure 14: Temperature dependence of the surface resistance of the coax cavity (upper) and field dependence of the Q-value at 2K (lower). The sample received different surface treatments [21].

A set of experiments conducted at Helmholtz Zentrum Berlin (HZB) investigated the flux trapping and flux release properties of polycrystalline, large grain and single crystal niobium subjected to different surface treatments as shown in Fig. 15 [61]. Magnetic flux could be trapped in the disc or rod-shaped samples by employing external magnetic fields during cool down and the amount of trapping was monitored with a flux gate magnetometer. The samples could be heated and the flux release could again be measured with the flux gate magnetometer. The experiments showed that 100% of flux was trapped in the fine grain (FG) material independent of cooling rate, whereas in the single crystal material annealed at 800 °C, the flux trapping depended logarithmically on cooling rate and saturated roughly between 60 and 70%. Since the mobility of flux lines is temperature dependent and increases towards lower temperature, the measurements indicate a significantly smaller pinning force in the annealed single crystal material, consistent with results obtained from magnetization measurements at JLab [25].

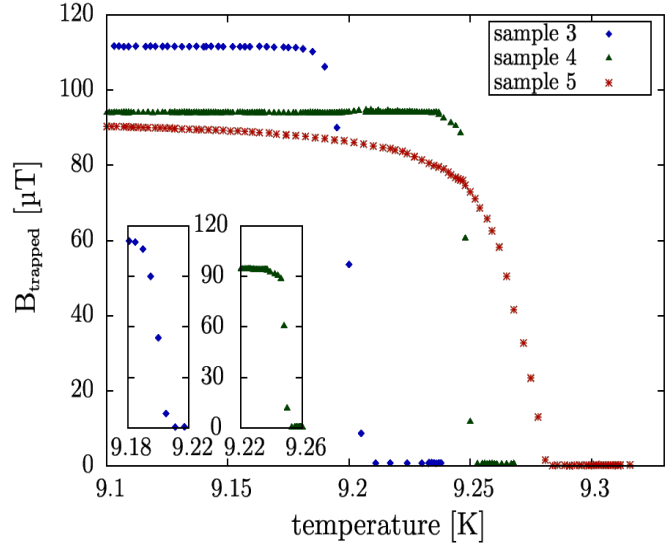

Figure 15: Flux release as a function of sample history: Sample 3 is FG, BCP and 800 °C HT for 3 hrs; sample 4: single crystal BCP, and sample 5: single crystal, BCP with 800 °C HT for 3 hrs [61].

### 2.8 Tunneling Measurements

It has been suggested that tunneling experiments could be used to investigate the electronic structure and density of states of niobium surfaces at different oxidation states. The reason for such investigations would be to clarify the importance of the niobium-oxide interface in causing the non-linearities in $Q_0$ vs $E_{acc}$ tests as suggested, for example, by the "interface tunnel exchange" model of ref. [62]. Tunneling spectroscopy is an ideal surface sensitive method to probe both the energy gap and the electronic density of states. The instrumentation for the point contact tunneling consists of a gold tip touching the niobium sample under test and the I-V curve and its derivatives $dI(V)/dV$ are taken. Such a measurement was done on an electropolished single crystal sample after air exposure and after low temperature baking in air at 120 °C for 48 hrs. The conductance measurement gives a gap parameter of $\Delta = 1.55$ meV indicative of bulk niobium, but also shows a smearing of the density of states, which was attributed to magnetic scattering caused by the magnetic moments of the non-stoichiometric surface oxides on the niobium. After baking, the broadening is somewhat reduced, hinting at the formation of a more stoichiometric oxide with reduced magnetic moments [63].

The point contact tunneling method was successfully used to find differences between so called "hot" spot samples and "cold" spot samples cut out from large grain niobium cavities [38], which had been tested with temperature mapping and had shown different power dissipation patterns on the cavity surface. The analysis of these measurements showed that in the "hot" samples, characterized by a lower energy gap and a larger pair breaking parameter as compared to the "cold" samples, superconductivity has degraded and more unpaired electrons add to the dissipation.

## 2.9 Field Emission

Enhanced field emission (EFE) is one of the limiting effects in superconducting accelerating cavities made from niobium. At high peak electric rf fields ($E_{peak} > 40$ MV/m) electrons are drawn out of the niobium surfaces or contaminating particulates and are accelerated across the accelerating gap in the cavities by the rf fields. When impacting on opposing surfaces they generated X-rays and additional heating of the surface, degrading the performance of the cavities. First experiments on niobium samples in a dc test set-up to explore the nature of field emission sites and to correlate the results with field emission in rf cavities were done at CERN in ~1985 [64]. Later on this type of investigations was championed by the University of Wuppertal and much of the knowledge about FE on niobium has been gained at this institution.
The results can be summarized as follows:
- Large grain or single crystal niobium samples show very smooth surfaces after an appropriate amount of material removal by BCP. After 100 μm removal the surfaces roughness is ≤10 μm; the surface is rougher after less material removal.
- The onset for field emission depends on the surface roughness; for very smooth single crystal samples the onset field was as high as 200 MV/m.
- As shown in Fig. 21, the emitter density and the onset level for EFE is clearly lower for large grain and single crystal niobium compared to the best electro polished fine grain niobium.
- No field emission from grain boundaries was detected up to electric fields of 250 MV/m.
- After low temperature bake at T ~ 150 °C some evidence for grain boundary assisted FE was reported.
- Good quality single crystal samples made it possible to measure the intrinsic field emission at selected defect-free local areas. In such areas on-set fields larger than 1 GV/m were found (see Fig. 20).
- Measurements of intrinsic field emission on crystals with different crystal orientations suggest an anisotropy in β-value between [100] and [111] orientation ($\beta$ is the field enhancement factor).

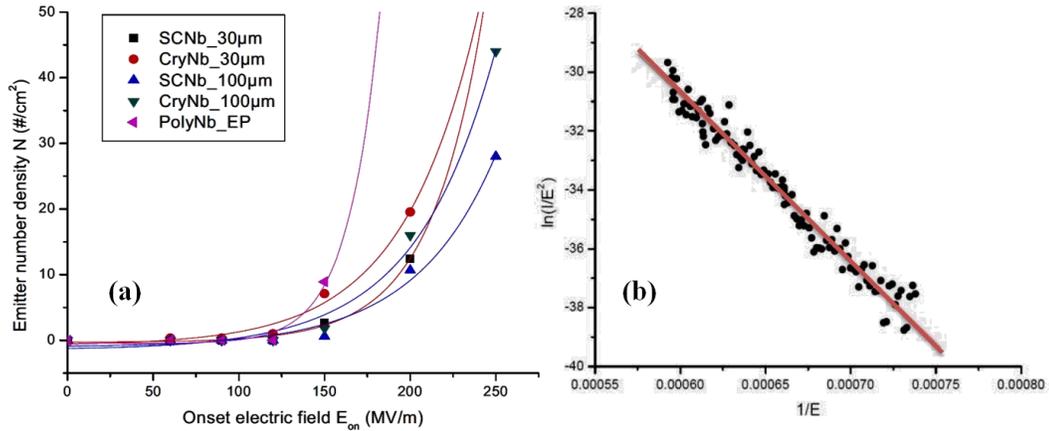

Figure 16: (a) Emitter density as a function of on-set field for EFE for differently treated niobium samples (b) Fowler-Nordheim plot of intrinsic field emission of single crystal niobium with an enhancement factor of β=1 and a work function of Φ=4 eV [65].

## 3. CAVITY FABRICATION

### 3.1 Slicing of ingot Nb

Besides the aspects of material formability, crystal structure and crystal deformation there is the technical aspect of fabricating cavities from this material in a reliable and cost effective way. This includes the melting of the ingot, the slicing of sheets, deep drawing and reproducibility of deep drawn parts, machining and electron beam welding. One of the first large grain ingots JLab received from CBMM had a very large single crystal in the center of the ingot as shown in Fig. 17(c). This is a very desirable configuration – for deep drawing of half cells the highest deformation occurs in the center and a single crystal in this area provides uniform mechanical properties – however, more often the ingot consists of several large grains. Unfortunately it was unknown under which conditions and configuration the ingot from which the slice shown in Fig. 17 (c) was grown and how it could be reproduced.

Several attempts have been made to grow single crystal ingots. Within the framework of an R&D program with DESY, W.C. Heraeus investigated the growth dynamic [66] in an attempt to produce reliably ingots with a large central grain and a specified orientation as required by the DESY specifications [67]. There are many variables in Electron Drip Melting, which is the only applicable process for industrial use schematically shown in Fig. 18, influencing the nucleation and dissolution of grains in the liquid pool such as local pool temperature, molten pool motion, external vibrations, dripping of melt-stock into the pool, vibration from the withdrawal of the ingot [68].

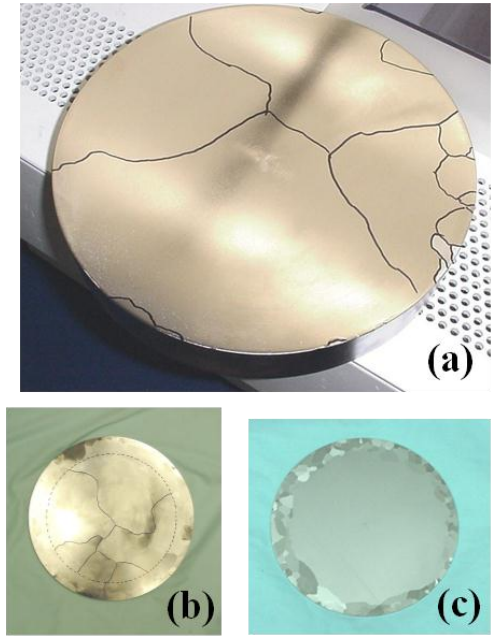

Figure 17: Large grain ingots from CBMM; (a) and (b) have a more typical grain structure, (c) has a large single crystal at the center.

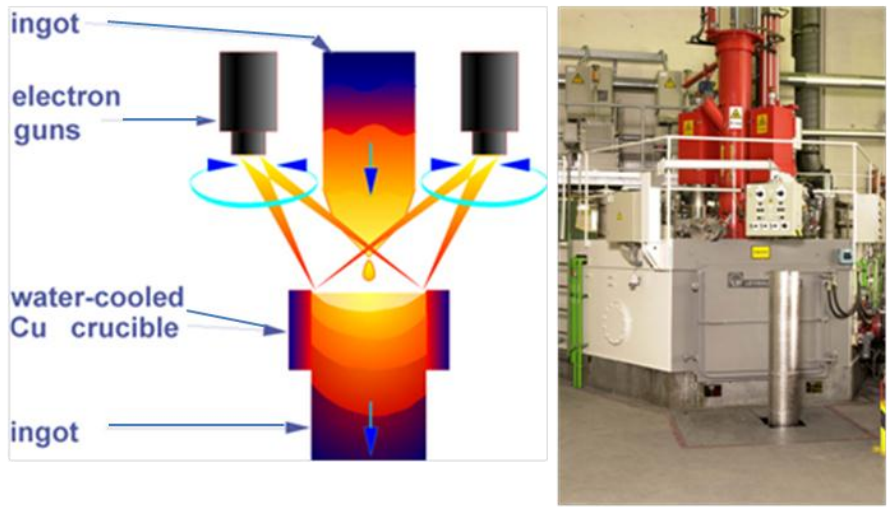

Figure 18: Schematic of drip melting in an Electron beam melting furnace (left) and furnace at W.C. Heraeus (right).

    The efforts at W.C Heraeus, which included variation of melting/cooling parameters such as modification of the beam shape, position and energy entry and cooling of the crucible side wall or the bottom, resulted in the realization that the melting process is not sufficiently stable to reproducibly create a central crystal of 150 mm diameter with a specified crystal orientation throughout a whole ingot of approximately 2000 mm length. However, the company succeeded in providing 250 large grain discs to

DESY from several ingots. Further efforts in growing single crystal ingots have been reported in ref. [69]; at Tokyo Denkai a seed crystal specimen was placed on a niobium bottom plate and the standard melting process was started. Ultrasonic tomography showed that the seed crystal was not sufficiently thick and that the base plate crystals determined the crystal growth in the ingot. It is concluded that a thicker seed crystal is needed for a successful single crystal growth.

A cost-effective slicing process was developed at W.C. Heraeus to provide sheets for deep drawing of cavity half cells from an ingot. The slicing procedure used the same multi-wire technique as it is used for silicon wafer slicing. This method was subsequently further developed at KEK involving Japanese industry for "mass production". Presently a powerful multi-wire slicing machine using very thin piano wire (0.16 mm) and a slurry of abrasive material continuously applied to the wires is installed at Tokyo Denkai with a capacity of slicing simultaneously 150 discs from an ingot of approximately 11" diameter in about 48 hrs with high accuracy and reproducibility. Figure 19 depict the slicing schematic and actual arrangements.

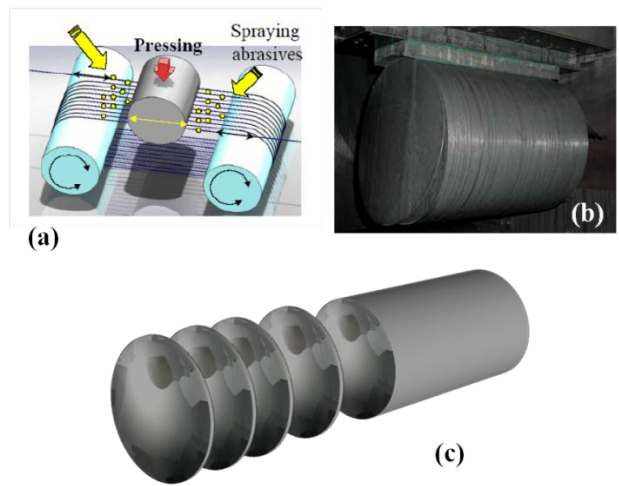

Figure 19: (a) Schematic of multi wire slicing, (b) actual slicing arrangement at TD (refs. 69, 70) and (c) W. C. Heraeus. The slicing is done with thin piano wire (0.16 mm) aided by a slurry of abrasive material.

Two aspects of the ingot slicing are important:
- The slicing provides less contaminated disc material with fewer defects compared to conventional sheet production, where contaminants from forging, rolling and annealing cannot be totally excluded. This observation can be drawn from eddy current scans performed at DESY on sliced discs.
- The sheets can be produced with very tight tolerances and very good surface finish, therefore mechanical grinding and buffing is not necessary.

## 3.2 Material costs

One of the promises for the ingot material was the expectation of lower material costs compared to sheet stock due to a much reduced handling and material waste caused by the rolling process. Figure 20 shows schematically the simplified production process [70]: instead of having 16 processes from starting material to final product in the conventional sheet production sequence, the number of handling procedures is reduced to 7. Additionally, the amount of material waste is much reduced in the case of ingot slicing as has been discussed in ref. [69]. It has been estimated that the cost of sliced ingot sheets in mass production could be as low as half the cost of conventional fine grain sheet material [70,71].

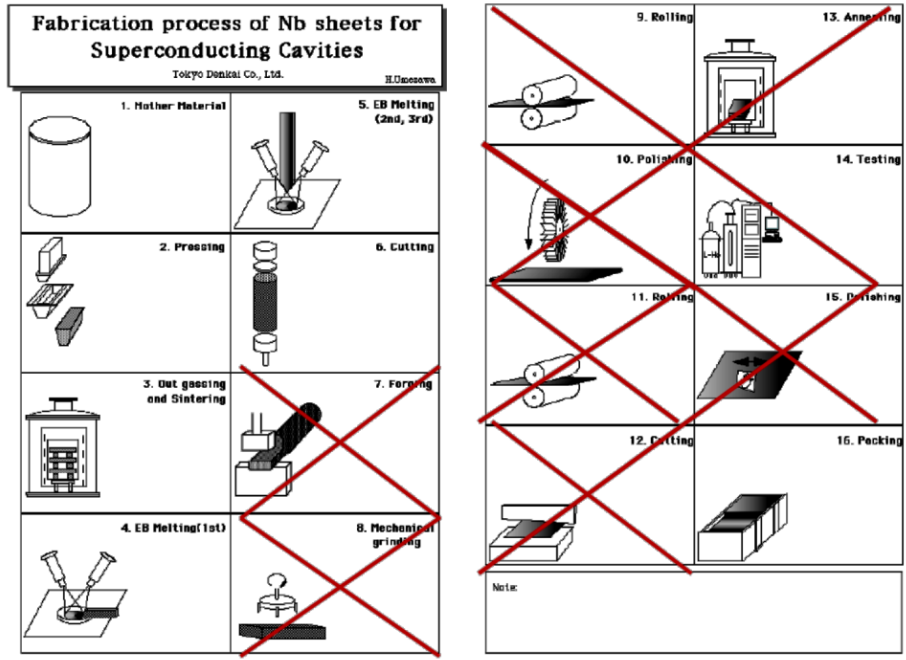

Figure 20: Schematic of the simplified production process for ingot niobium sheet [70].

Obviously the cost for the starting material is an important contribution to overall material cost. Presently the specifications for RRR niobium used for SRF cavities are quite stringent for impurity contents. One of the main cost drivers is the Ta contents: because Ta is found with Nb in the ore and the separation of both materials of similar chemical nature is cost intensive. It has been proposed to relax the specifications for Ta contents based on experimental evidence with cavities built from higher Ta contents niobium [72], which reached performance levels comparable to cavities fabricated from lower Ta contents material. Magnetization studies on small samples cut from ingot Nb with Ta content between 150 and 1300 wt.ppm did not show any significant difference in the superconducting properties such as $T_c$, $H_{ffp}$ and $H_{c2}$ of the various samples, in agreement with the SRF cavity study [73]. Nb with higher Ta content than typically specified for fine-grain Nb ($\leq$ 500 wt. ppm) is available at lower cost.

### 3.3 Cavity Fabrication

The first single cell 1500 MHz cavity of the "High Gradient" shape was fabricated in 2004 at Jefferson Lab from wire EDM slices from CBMM ingot niobium using the same fabrication techniques developed for fine grain material. Even though the half cells for this cavity and the one's following afterwards showed ragged edges as shown in Fig. 21 and other manufacturing problems such as thinning or ripping in the iris region when grain boundaries came together, spring back and "oval" shaping due to different crystal orientations and slippage of grain at the grain boundaries, none of these difficulties were "show stoppers". This was also reported from more experienced cavity manufacturers such as ACCEL Instruments (now Research Instruments). ACCEL Instruments [74] manufactured by 2006 in total 4 single cell cavities and 3 complete 1.3 GHz 9-cells TESLA cavities. The half cells were deep drawn using a stamp and a cushion; non-uniformities were seen in the area of the equator. In addition steps at grain boundaries as high as 0.2 mm were reported. These grain boundary steps could be smoothened out by mechanical grinding. The grain boundary steps were less pronounced when spinning techniques were used for the forming of the half cells.

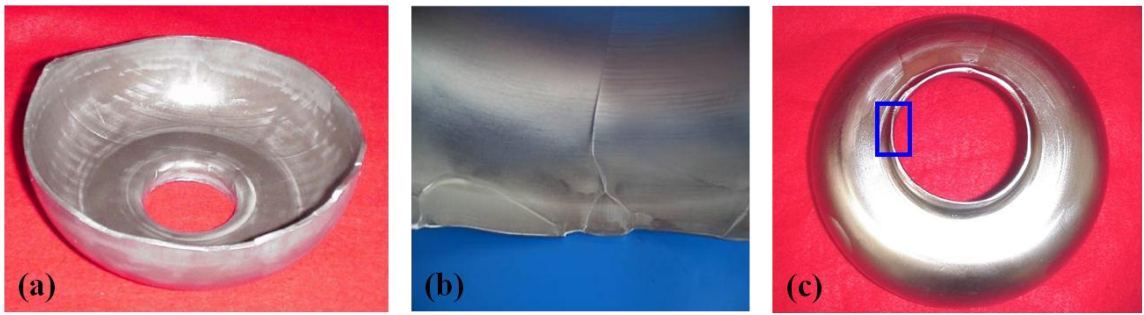

Figure 21: Half cells deep drawn from large grain niobium discs showing (a) earing, (b) grain boundary slippage and (c) thinning as indicated.

The deviation from roundness of the half cells caused some more elaborate assembly procedure for electron beam welding; during the welding itself no differences were observed. No manufacturing problems were detected during the production of the 9-cell cavities and the correct cavity frequency was achieved by frequency control of the half cells prior to welding as in standard manufacturing. The deep drawing behavior of half cells made from ingot material was analyzed at DESY [75] with following conclusions:
- Frequency deviations and standard deviation for the center half cells of a 9-cells cavity are smaller than the same parts deep drawn from fine grain material.
- The deep drawing behavior of this material is different than for fine grain material especially the spring back is different but it is more stable and allows producing more uniform half cells.
- Large frequency deviations can be reduced by modification of the deep drawing tools.

These early experiences with the manufacturing of single cell and multi-cell cavities answered several concerns about the ability to fabricate cavities from this material as listed above:
- Grain boundaries are leak tight even at cryogenic temperatures.
- Electron beam welding causes no difficulties, even across pronounced grain boundaries.
- Deep drawing operations produce uniform half cells.
- Different grain orientations have different mechanical response to deep drawing causing e.g. slippage at grain boundaries; however, such steps at grain boundaries can be managed by mechanical grinding and are no "show stoppers".
- Even though a large central crystal in a niobium sheet is preferable for the deep drawing, deep drawing of sheets with randomly distributed large grains is successful.
- However, as the experience at JLab has shown, sheets with large amount of stresses as in machined sheets will fail in the deep drawing process. In this case a stress relieving heat treatment at 600 °C for 10 hrs solved the difficulty.

One of the first ingots JLab received from CBMM had a large central crystal. However, this crystal was not large enough for a 1500 MHz CEBAF type cavity; therefore at Jlab two higher frequency cavities (2.2 GHz and 2.3 GHz) of two different shapes were fabricated to explore the rf performance of single crystal niobium. As expected, the fabrication process was without problems; no oval-shape appearance of the half cells and less spring back and the performance of the cavities were excellent as discussed below.

The single crystal option was further pursued at DESY and later at PKU based on the DESY experience described here with the goal to produce cavities of the same frequency as the large grain cavities (1300 MHz). In order to accomplish this, a procedure had to be developed to increase the size of a smaller single crystal to the size needed for a half cell without changing the crystal orientation and subsequently prevent recrystallization at the electron beam weld. The starting material is a large grain disc provided by W.C. Heraeus with a large center crystal of [100] orientation and approximately 200 mm diameter. To achieve the sheet size for a half cell a 50% deformation by cross rolling was needed.

A detailed description of the single crystal enlargement and subsequent electron beam welding procedures is given in ref. [17]. A condensed version is:
- Crystal orientations [100] and [110] have successfully been enlarged with deformations up to 50% and subsequent annealing between 800 °C and 1200 °C as confirmed with crystallographic investigations on TEM samples and with SEM pictures of etch pits.
- Crystals of [111] orientation are recrystallizing after deformation and annealing at 1200 °C.
- During electron beam welding two single crystals grow into one single crystal if the crystallographic orientations are matched at the seam. Unmatched orientations produce a pronounced seam.

Several single crystal cavities were fabricated in this manner and subsequent performance tests will be described in next section.

## 4. Cavity Test Results

The advent of ingot niobium for SRF cavities had attracted the interest of many institutions and work on fabrication and testing of such cavities was carried out initially at Jefferson Lab, DESY, KEK, Michigan State University, Cornell University, Peking University, Institute of High Energy Physics China and BARC India. The work on large grain niobium at Fermi Lab and Brookhaven National Lab were mentioned in few conference presentations, however no published results were available at the time this manuscript preparation. In the following the activities in each laboratory will be reviewed.

### 4.1 Jefferson Lab, USA

The evaluation of ingot niobium for SRF cavity application started at Jefferson Lab in 2004 as a collaboration with CBMM. CBMM provided 2 ingots named "A" and "B" of approximately 220 lbs each with a diameter of 9.25". Ingot A had a very large single crystal of 7" diameter in its center with a few grains at its periphery, whereas ingot B had several large grains, but not as large as ingot A. The RRR value of the material was ~ 280 with a Ta content of ~ 800 ppm. Post-purification at 1250 °C in the presence of Ti did not significantly improve the RRR value (~ 10%). As discussed later, the large single crystal of ingot "A" was used to fabricate a 2.3 GHz single cell cavity. In this phase of exploring the capabilities of large grain ingot material, material from four different manufacturers, as listed in Table 2, was used to fabricate single cell cavities, which subsequently were subjected to the same standard surface treatments such BCP, hydrogen degassing, BCP, high pressure rinsing (HPR) and, "in situ" LTB. Even though the sheets for the half cells were produced in different ways (saw cutting, wire EDM) and the material had different grain sizes, RRR values and Ta content, the performances after post-purification are astonishingly very similar as seen in Fig. 22.

Table 2: Properties of ingot niobium supplied by different companies

| Manufacturer | Ta content (ppm) | RRR | Cutting method |
|---|---|---|---|
| CBMM | 800-1500 | 280 | Wire EDM |
| Ningxia | < 150 | 330 | Saw cutting |
| W. C. Heraeus | <500 | 500 | Wire saw |
| Wah Chang | <500 | > 300 | Saw cutting |

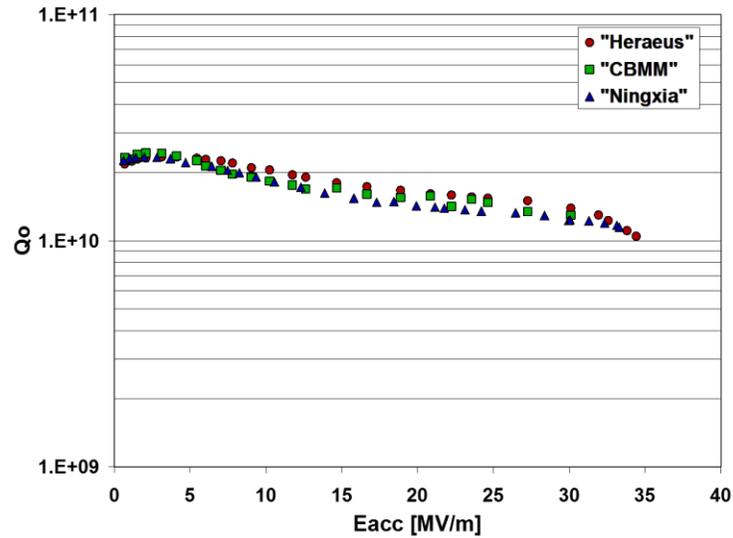

Figure 22: $Q_0$ vs $E_{acc}$ at 2K for single cell cavities made from material of 3 different vendors after post purification at 1200 °C [32].

The comparison of material from different vendors was continued to get better statistics. Five cavities each were fabricated from CBMM, Tokyo Denkai, Ningxia and W.C. Heraeus and all cavities were subjected to the same surface treatments. This test series started in 2007 with the "standard" surface treatment and initial results have been reported in [76]. Figure 23 displays the best results obtained from W.C. Hereaus, Ningxia and Tokyo Denkai cavities. In some cases after the first assembly and test the performance as shown has been reached; in other tests an additional high pressure water rinse (HPR) was necessary to eliminate some field emission. Except for TD #1 all results shown are after LTB, which eliminated the "Q–drop" (Q degradation at high field). From the set of the CBMM cavities only one cavity was tested with a side-port welded to one half cells at approximately 45 degrees from the equator. This cavity quenched at $E_{acc}$ ~ 29 MV/m. The remaining CBMM cavities need to go through the testing procedure; also all 20 cavities need to be re-testing after a post-purification heat treatment to complete the "statistical" evaluation. This will be done, when the resources are available. For a direct comparison of the material performances one has to look at the magnetic surface fields, at which the quenches in the cavities occurred. Since the cavity from W.C. Heraeus is of the low loss variety (LL), favoring lower magnetic fields for the same accelerating fields as in the ILC-type cavities, the highest gradient of $E_{acc}$ ~ 45 MV/m in the LL cavity corresponds to a peak magnetic field of $H_{peak}$ ~ 160 mT, equivalent to the 35 MV/m was reached in the ILC-type cavities (within 10%).

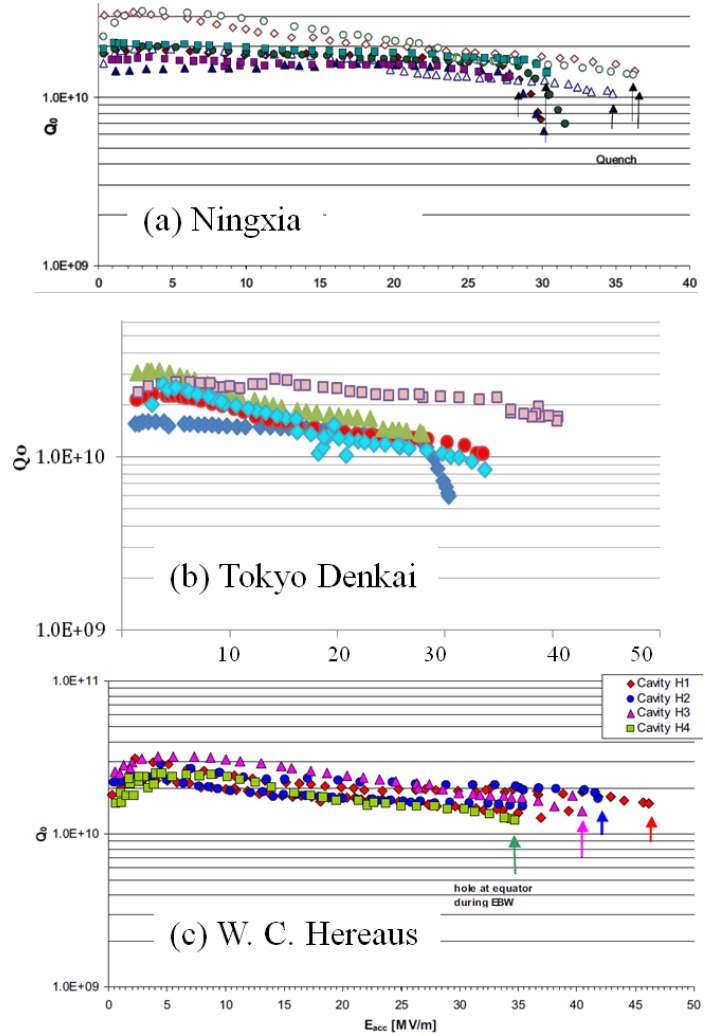

**Figure 23:** $Q_0 (E_{acc})$ measured at 2.0 K on cavities of different shape made from ingot niobium supplied by 3 different companies ( W.C.Hereaus, Ningxia, Tokyo Denkai) [76]

A series of experiments were conducted with cavities made from material with different grain sizes in an attempt to explore the influence of grain boundaries on limitations in SRF cavities. Four cavities of identical shape resonating at 2.2 GHz made from high purity niobium with single crystals, large grains of the order $cm^2$ size (2 cavities) and standard polycrystalline niobium were investigated [77]. After several steps of identical procedures (for material removal only BCP was used) the cavities showed very similar behavior after a final step of post-purification and LTB (see Fig. 24) with a slightly higher quench field for the single crystal material. This result does not point to any significant influence of grain boundaries in limiting the performance of these cavities.

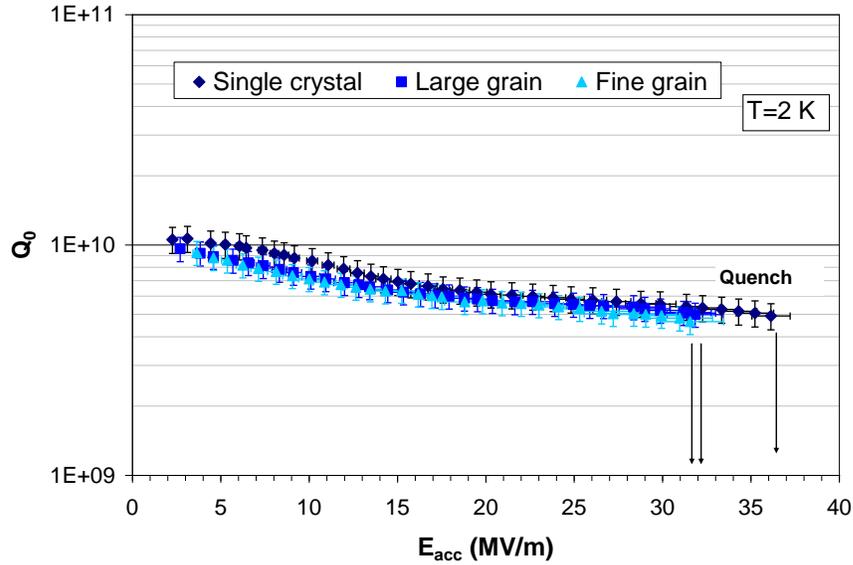

Figure 24: $Q_0$ vs. $E_{acc}$ at 2 K for three 2.2 GHz niobium cavities with different grain sizes and Ta contents. The single crystals has Ta content of ~ 800 wt. ppm and large grain and fine grain with Ta < 500 wt. ppm [77].

The cost for a useable cavity is determined by the material cost, the fabrication and the surface treatment. Material cost is determined among other factors by its purity and Ta content is a main cost driver; since Ta is substitutional impurity. It does not compromise the thermal conductivity (RRR value), which is believed to be important for high gradients. Therefore a set of experiments had been done in 1999 [72] with poly-crystalline niobium of different Ta contents without showing any significant differences in performance. Therefore, also large grain cavities have been evaluated made from higher Ta content sheets. These are exclusively ingot sheets from CBMM material. As shown in Fig. 24, despite significant differences in Ta content the performance of the cavities was quite similar.

More cavities from CBMM material with higher Ta content showed performances of $E_{acc}$ > 30 MV/m (or $B_{peak}$ > 130 mT) such as a 9-cell TESLA cavity or a 7-cell upgrade cavity J-100, with $H_{peak} \geq 110$ mT. Unfortunately, the multi-cell cavities fabricated at JLab had in nearly every case a hole in one of the cell's the equator weld. Even though these holes could be fixed satisfactorily, in several cases these areas limited the cavity performance. Although the causes for this difficulty during electron beam welding are unknown, clues suggest less than ideal cleaning of the parts prior to welding as a possibility. The performance of 1.3 GHz multi-cell cavities built at JLab from large-grain Nb is shown in Fig. 25 [78].

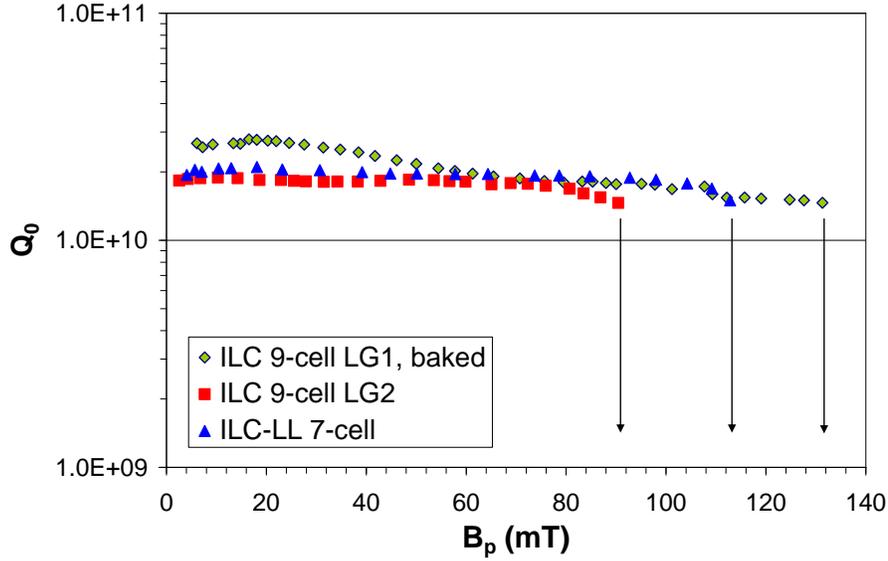

Figure 25: Best rf performance at 2 K for the three 1.3 GHz multi-cell cavities built at JLab [78].

A large-grain cavity built from CBMM Nb was used to study the impact of trapped magnetic flux on rf losses at medium field and to image the surface resistance by low temperature laser scanning microscopy (LTLSM) [79]. LTLSM is a successful method to characterize thin films of high $T_c$ material in microwave resonators [80]. With a focused laser beam the surface under investigation is scanned point-by-point and the photo-response induced by the interaction of laser light with the superconductor is recorded as a function of the laser spot position in (x,y). The same method can be used to map out the surface resistance of a niobium cavity. A special cavity operating in the $TE_{011}$ mode at 3.3 GHz as shown in Fig. 28(d) was used for the measurement. The resulting maps, which have spatial resolution of ~2 mm, show non-uniformity of the surface resistance, correlated with "hot spots" by temperature mapping of the outer surface. If the losses in these areas are caused by pinned fluxoids, then they should move or disappear, if they can be de-pinned by applying sufficient heating by the laser to overcome the pinning force. Sweeping the surface with the laser at higher powers showed that the T-map obtained afterwards had changed and hot spots had moved, possibly indicating a re-distribution of fluxoids after they had been "de-pinned" from their original location with the laser power [81].

By taking laser scans at different laser power levels it is possible to obtain information about the temperature dependence of the surface resistance at different areas on the surface. Analysis of the data with Halbritter's surface resistance program [82] show that at the "hot spots" from the T-map weaker superconductivity (smaller gap value) is present. The causes for these losses are not well known, although the changes in the "hot spot" distribution after laser "sweeping", suggest that pinned fluxoids contribute to non-linearities in the rf surface resistance.

The lossy areas in a cavity surface as described above are responsible for two features, namely a residual resistance preventing the cavities from reaching the theoretical Q-values at the operating temperature, in most accelerator application at frequencies above 500 MHz being 2K or 1.8K, and a decrease of Q–value with increasing cavity gradient. For CW application of an accelerator it is desirable to have Q-values as high as possible. For a typically accelerating cavity of the CEBAF variety a Q-value of $Q(2K) \sim 4 \times 10^{10}$ at accelerating gradient in a range of 15 MV/m $\leq E_{acc} \leq$ 25 MV/m is theoretically possible, but typically only 1/3 of this value is achieved routinely, when standard processing steps such as EP, hydrogen degassing and high pressure rinsing are applied. A significant effort has been made at JLab to achieve such a goal.

For a long time one of the authors of this review (G.R.M) has suspected that hydrogen, which is readily absorbed by niobium if a protective oxide layer is missing, is a major contributor to the residual resistance and the non-linearities in $Q$ vs $E_{acc}$. The idea therefore emerged to prevent the hydrogen re-entering into the niobium after heat treatments at elevated temperatures (> 600 °C) at which hydrogen was degassed from the niobium. Typically, after a hydrogen degassing treatment, a cavity is chemically etched to remove ~15–20 μm from the surface, which has been contaminated during the high temperature treatment and cool down by the residual gas species in the furnace. During this chemical treatment niobium re-absorbs again a large amount of hydrogen from the initially low concentration after the heat treatment as measured on test samples. Therefore, the step of chemical treatment after furnace heating was eliminated and only ultrasonic cleaning and high pressure rinsing with ultrapure water was applied to the cavities and prior to cryogenic testing the cavity was baked "in situ" for 12 hrs at 120 °C. In a scan of varying the heat treatment temperature between 600 °C and 1200 °C an improvement of 30 % of the Q-value at an accelerating gradient of ~20 MV/m was achieved at temperatures between 800 °C and 1000 °C [83]. However, at higher temperature, contamination of Ti from the furnace occurred and no further improvements could be achieved. It became obvious that the cleanliness of the furnace seemed to be important. Therefore a dedicated furnace of the induction type [84, 85] was commissioned and the studies continued. The furnace implemented at JLab and it consists of a quartz bell jar with an inner niobium chamber, which houses the cavity. The cavity vacuum system is separated from the bell jar vacuum. The niobium chamber is induction-heated by the copper coil and the cavity is heated by radiation. The vacuum in the niobium chamber at the temperatures of interest is better than $10^{-6}$ torr.

In this new furnace the heat treatment process was modified in 2 ways: firstly, after the initial degassing the furnace was purged with argon gas at a pressure of $\sim 10^{-6}$ torr to prevent re-contamination of the niobium from residual gas in the chamber and secondly, once the cavity was cooled down, the furnace was vented with high purity oxygen to seal the cavity surface with a protective oxygen layer against hydrogen pick-up [86]. As discussed in ref. [87] such a "dry" oxide layers will be beneficial, because it is forming with less defects than a "wet" oxide. This procedure was investigated in the temperature range between 600 °C and 1400 °C; after each test there was a low temperature bake test and subsequently a new baseline was established by removing app. 30 μm by BCP followed by a baseline test [88]. This procedure resulted in an improved $Q$–value at medium cavity gradient ($B_{peak}$ = 90 mT corresponding to $E_{acc} \sim$ 21 MV/m ), culminating in an exceptionally high value after heat treatment at 1400 °C as shown in Fig. 26. Also shown in the figure is also the T-dependence of the surface resistance [89].

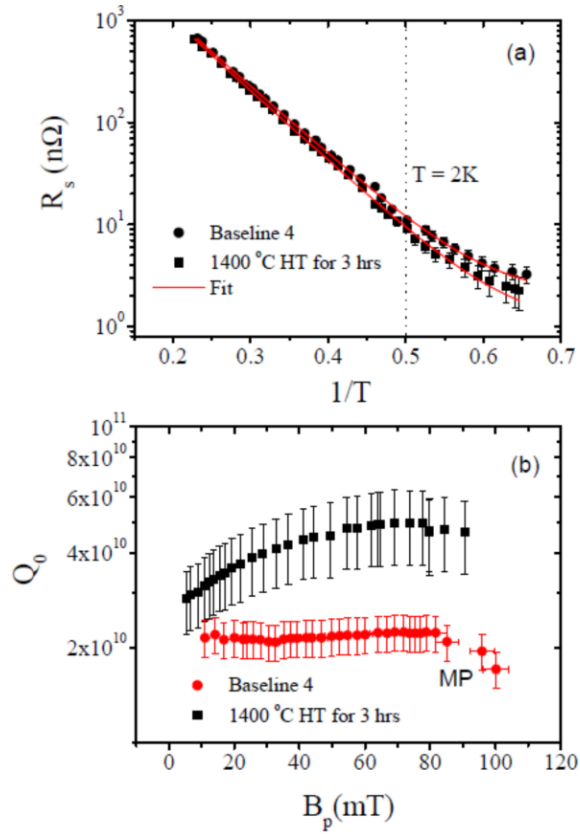

Figure 26: Improvement of cavity performance after heat treatment at 1400 °C for 3 hrs, (a) $R_s$ as a function of temperature and (b) $Q_0$ vs $B_p$ at 2K [89].

The increase in $Q$-value with increasing field is somewhat unusual and not yet fully understood. Further studies on the reproducibility of the high $Q_0$-value and to understand the heat treatment effect are planned

A series of tests was carried out with successively removing small amounts of material from the surface and to determine the evolution of the cavity performance as a function of damage layer removal. This series was done with a single crystal cavity of the Tesla shape provided by DESY. Surface layer of approximately 112 μm from the inner surface and $E_{acc}$ = 37.5 MV/m after only 6 hrs of LTB at 120 °C which improved the cavity performance significantly as shown in Fig. 27. Further steps of material removal resulted only in a slight improvements of the breakdown field.

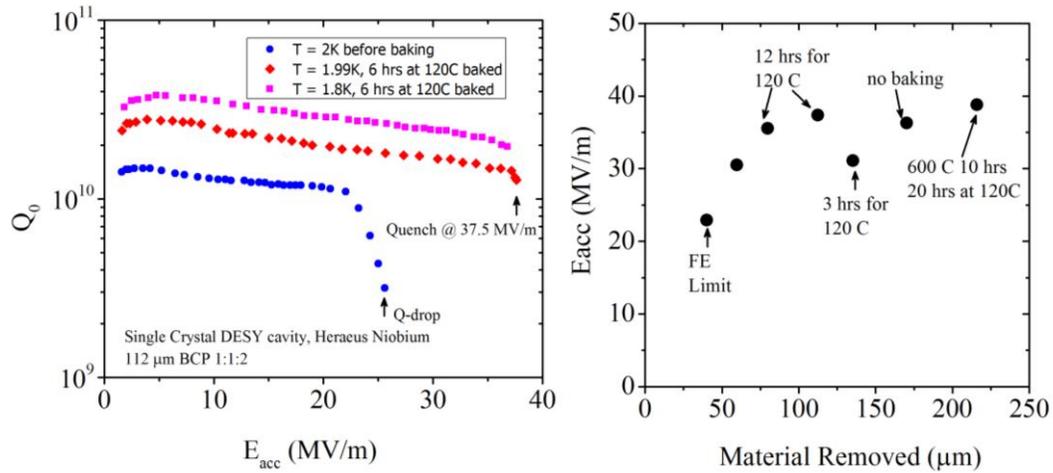

Figure 27: Effect of material removal on cavity performance; this test series was done on a single crystal cavity fabricated at DESY [90].

As of today, JLab has fabricated around 50 single cell cavities "in house" from material of different origins; these cavities have been evaluated in more than 200 tests. Additionally, four 9-cell cavities each of the TESLA and ICHIRO shape, two 7-cell upgrade cavities, one 7-cell LL cavity and one high current 5-cell cavity have been built. "Unconventional" cavity shapes such as a crab cavity, a 3.5 cell photo-injector cavity, several 0.5 cell cavities and 1.5 cell photo-injector cavities complete the JLab "inventory". Figure 28 shows several of these cavities.

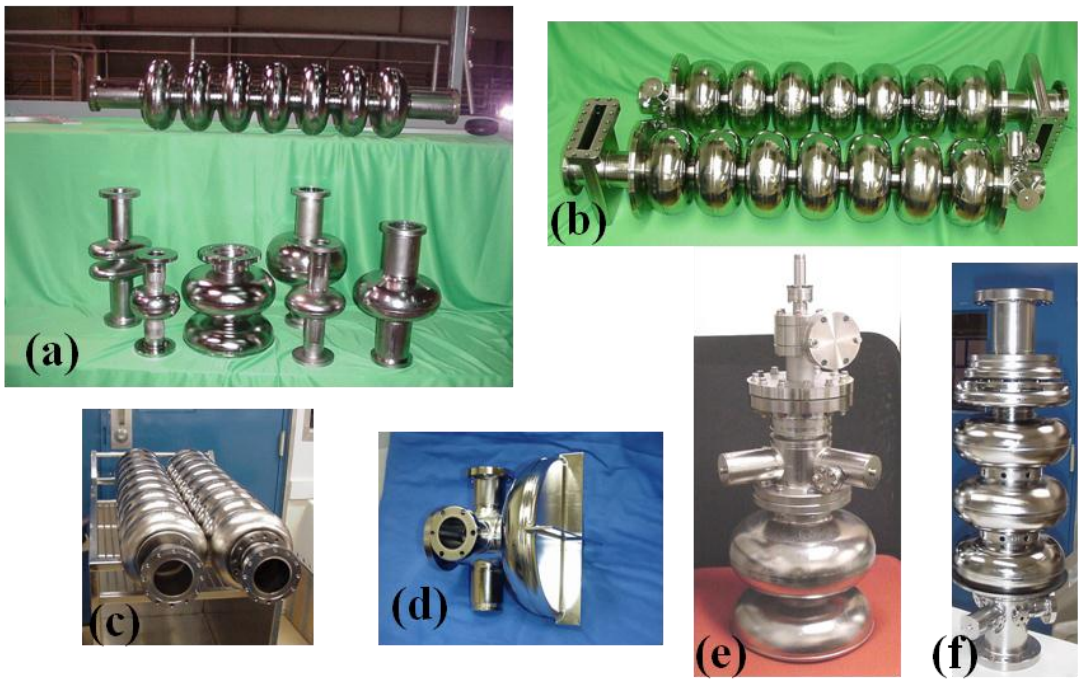

Figure 28: Collection of cavities (single and multi-cell) fabricated at JLab from large grain/single crystal material. (a) 7-cell cavity, 2-cell spoke, a single crystal 2.3 GHz LL, a 1.5 cell cavity for joint tests, a 2.3 GHz TESLA shape single crystal cavity, (b) and 2

LL and TESLA single cells upgrade cavities, (c) 2 Ichiro-type 9- cell cavities, (d) a half-cell for laser heating experiments, (e) a 1.5 cell photo-injector and (f) a 3.5 cell photo injector cavity of the Rossendorf design.

The JLab experience with large grain and single crystal cavities can be summarized as follows:
- In most cases no major problems with fabrication were encountered, even with "unconventional" shapes. An exception was material from Ningxia, which had to be stress relieved (600 °C, 10 hrs) to deform well.
- Nearly all cavities build "in house" were only surface treated by BCP, resulting in smooth surfaces.
- Material from different supplier with different RRR – values and impurity concentrations – especially high Ta content – performed very similar.
- No differences were found in cavity performance on the method by which the sheets for half cell were obtained: wire EDM or saw cutting
- A comparison between cavities made from material with different grain sizes did not show significant differences – the single crystal cavity had slightly higher breakdown fields.
- In many experiments the best procedure for "in-situ" baking of BCP cavities was explored, in order to eliminate the "Q–drop". The experiments showed that a temperature of 120 °C and durations between 6 hrs and 24 hr are successful. This needs to be compared to electropolished fine grain material, where baking times of 48 hrs are the norm.
- For best performance, approximately 100 μm of material have to be removed from the surface, when conventional procedures such as BCP and hydrogen degassing are used. The single crystal cavity made at DESY reached a gradient of $E_{acc} \sim 37$ MV/m after a removal of 112 μm BCP and 6 hrs of baking at 120 °C.
- Cavities from large grain ingot material show higher onset values for the Q–drop compared to poly-crystalline niobium.
- Development of Q-improvement procedures was successfully carried out with a large grain cavity. The Q–value at 2K improved by a factor of 4 to $\sim 4.6 \times 10^{10}$.
- Multi-cell cavities made from ingot material did not perform as well as single cells; however, in many cases unexpected and not-understood problems during electron beam welding occurred. Such problems do not seem to be related to the material but rather to the need for more stringent quality control of the cleaning of the Nb assemblies prior to weld.

### 4.2    DESY, Germany

The activities and accomplishments at DESY have been summarized in a recent publication [91]. For completion of this review only the highlights of the work at DESY are reported in the following.

In contrast to Jefferson Lab, which made use of its own fabrication and rapid prototyping capabilities for exploring ingot niobium, DESY collaborated closely with industry and many of the developments reported below were the results of this close collaboration. The R&D program initiated at DESY in 2004 pursued two aspects:
- To explore fabrication and preparation procedures and
- To gain a basic understanding of the differences between large grain/single crystal and fine grain material and to analyze the potential of ingot material for X-FEL application.

The effort at DESY driven by the X-FEL project started with the development of ingot material with a central grain at W.C. Heraeus as discussed above. Simultaneously, a cost effective cutting procedure for discs to maintain the purity and surface quality long with tight thickness tolerances was developed at W. C. Heraeus. Single cell and multi-cell assemblies were fabricated at ACCEL (now RI) using material from W.C.Hereaus, Ningxia and CBMM. Along with the cavity fabrication material studies such as crystallographic structure investigations, measurement of mechanical and thermal properties and comparisons of surface treatments such as BCP, EP, heat treatment and baking at low temperature were conducted. Since no supplier of ingot material could provide a single crystal large enough to fabricate a single crystal cavity for the X-FEL project, the expansion of a smaller single crystal into a large disc useable for forming 1300 MHz half cells was pursued as discussed above.

In total eleven 9-cell X-FEL cavities were evaluated; the best of these cavities reached a final performance of $E_{acc}$ = 45.4 MV/m at a $Q$-value of $1.3 \times 10^{10}$, outperforming even the highest gradient cavities from poly-crystalline niobium. This accelerating gradient corresponds to surface magnetic field of $B_{peak}$=192 mT. Analysis of mode measurements of the nine pass-band modes showed that in some cells peak surface fields as high as $B_{peak}$ = 213 mT were reached, which is close to the fundamental limit of bulk niobium. Remarkably, the cavities made from large grain material exhibited higher Q-values at the operations gradient of the X-FEL as shown in Figure 29, which also shows the result from the best performing 9-cell cavity.

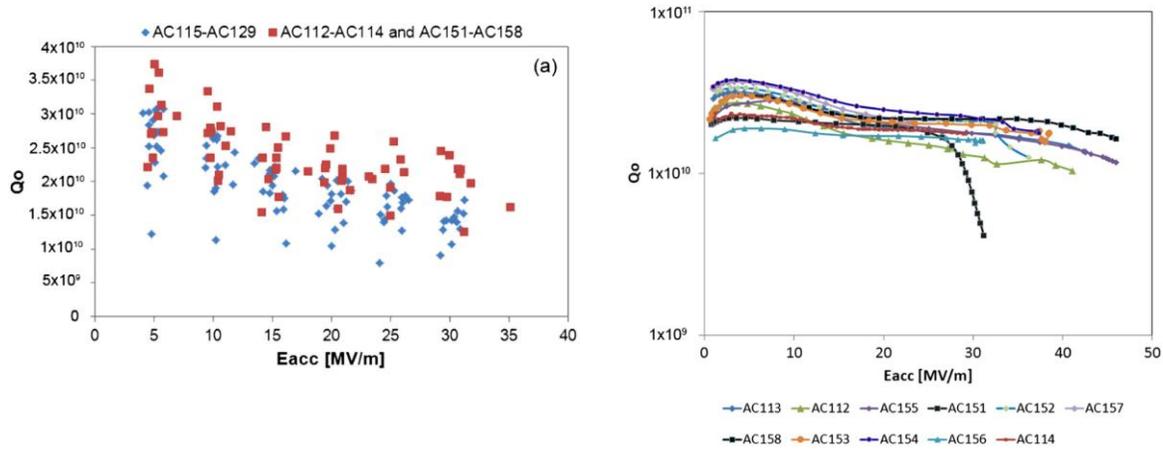

Figure 29: (a) $Q$–values as a function of $E_{acc}$ for LG (b) $Q(E_{acc})$ for 9-cell cavities AC112–AC114 AC151–AC158 at 2K. Notice the high $Q$-values [91].

The DESY experience can be summarized as follows;:

- Fabrication issues with shape accuracy, spring back and reproducibility could be solved by modifications of the dies. However, the crystal orientation of the center crystal of the disc has a pronounced effect on the shape accuracy of the half cells. Crystals with [100] orientation exhibited larger deviations from the ideal cell shape than crystals with [211] or [221] orientation. Nevertheless, the deep drawing behavior is more stable compared to fine grain niobium and allows production of more uniform cells.
- In a comparison between cavities made by deep drawing or by spinning the deep drawn cavities performed superior.
- EP as a final surface treatment produced higher cavity performances than BCP; in a test series, in which an EP cavity was BCP and subsequently EP again, the BCP treated surface had a poorer performance than both EP surfaces.
- Large crystals could be enlarged to the size needed for TESLA-type cavities and by maintaining the crystal orientation at the equator weld, the whole cavity remained a single crystal after electron beam welding
- 9-cell TESLA cavities performed very well after BCP treatment only and exceeded in the first phase of the R&D program the design values for the X-FEL in both Q-value and accelerating gradient by a large margin.
- EP surface treatment improved the cavity performance further and in two cavities record gradients of $E_{acc}$ ~ 45 MV/m were measured
- Two cavities have been installed in cryomodules and are operating in the FLASH accelerator without problems.

### 4.3 High Energy Accelerator Research Organization KEK, Japan

The excitement about ingot material as an alternative material for rf cavity fabrication and use spilled immediately over to Asia and by 2006 many activities had developed into R&D programs at KEK, IHEP and Peking University to investigate ingot material from different suppliers. At KEK the initial efforts with ingot material complemented the high gradient research on poly-crystalline niobium, which had been pursued for several years in the context of the proposed ILC requirements. This program had focused on developing the Ichiro-type cavity, which differed from the TESLA-type cavity shape by reducing the ratio of the peak magnetic surface field to the accelerating gradient by roughly 15%; this in turn will permit higher accelerating gradients at the same limiting magnetic field as given by the superconducting properties of the material.

The first two single cell cavities were fabricated from CBMM ingot material, followed by three cavities made from Ningxia large grain niobium. Similar fabrication difficulties such as slippage of grain boundaries and non-uniformity in the half cell shapes as in other laboratories were encountered during the fabrication, but did not prevent the completion and testing of the cavities. The same surface preparation steps as developed for fine grain material were applied. The typical process includes the CBP of ~200 μm, 800 °C for 3 hrs, light BCP or EP removing ~ 50 μm prior to the cryogenic test. The very encouraging result for the first cavity are shown in Fig. 30 together with data from a poly-crystalline cavity, which had been fabricated and treated in parallel with the cavity from ingot material: the large grain cavity had a very low residual resistance of $R_{res}$ = 1.46 nΩ; five times lower than measured on the fine grain cavity with accelerating gradient of $E_{acc}$ = 36.5 MV/m before it quenched [92].

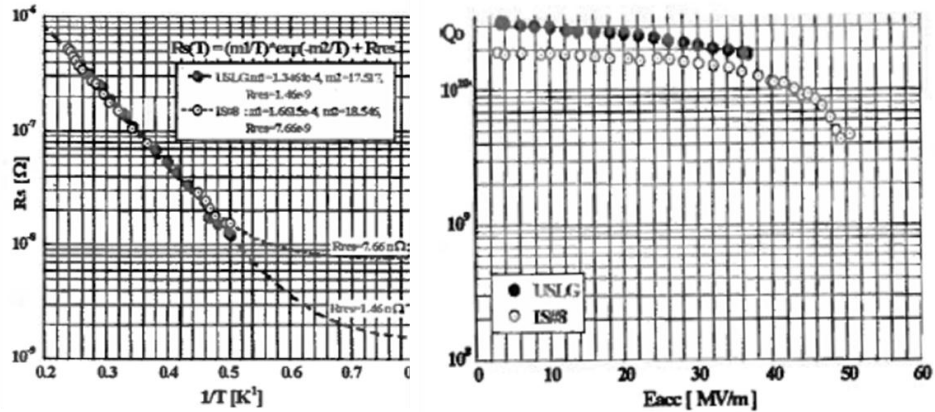

Figure 30: Performance of the first large grain single cell cavity fabricated at KEK from CBMM ingot niobium [91].

The three cavities made from Ningxia ingot material were used to investigate several issues:
- How do these cavities perform in comparison to fine grain cavities using KEK standard recipe?
- What is the effect of shorter "in situ" baking times on eliminating Q-drop?
- How does the cavity performance evolve with successive material removal steps by EP after CBP?
- How does the cavity performance evolve with successive EP material removal steps without prior CBP?

This study had the following results:
- The cavities perform very similar to cavities fabricated from fine grain niobium
- A baking time of 12 hrs is short enough to eliminate the *Q*-drop.
- A total of ~ 100 μm of EP is necessary for high performance in both *Q*-value and accelerating gradient.

Centrifugal barrel polishing (~ 200 μm material removal) seems to be essential to achieve high cavity performance; this is mainly due to removing weld irregularities, which have a degrading influence on performance.

A summary of results from the R&D with the cavities from Ningxia is given in Fig. 31: Both cavities with a CBP step performed very well, reaching gradients of 44 and 48 MV/m, whereas cavity #3 without CBP showed an early quench limitation.

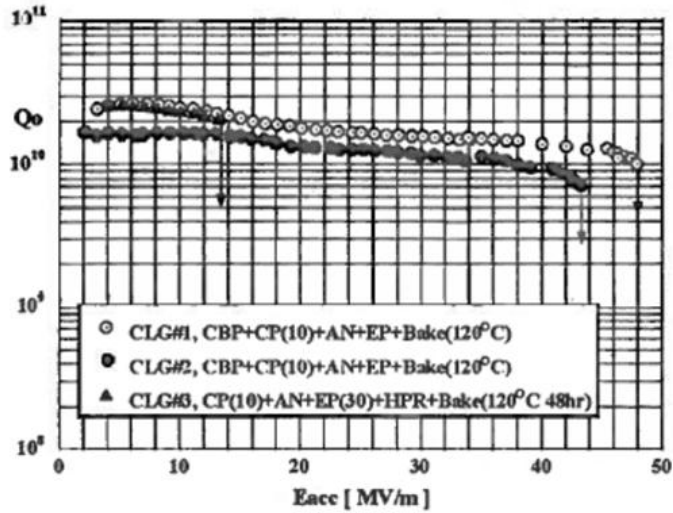

Figure 31: Summary of the performance of large grain cavities from Ningxia material. Note: Cavity CLG #3 did not receive CBP. The nomenclature used in figure CBP, CP (chemical polishing), AN (annealing), EP, HPR (high pressure water rinsing), and baking (120 °C baking) [91].

Subsequent work at KEK concentrated on evaluating ingot material from Tokyo Denkai - the company started production of large grain ingots in 2007 - and on developing in collaboration with industry a multi wire slicing capability based on machines used in the semiconductor industry for wafer slicing. This development has been discussed in earlier sections. The single cell developments pursued streamlining of procedures such as the use of BCP instead of EP and investigating the influence of modified procedures such as horizontal BCP. Additionally, reproducibility tests were conducted by repeatedly applying the same surface treatments alternating between horizontal and vertical BCP. In between a test series, the surface was also reset by an additional CBP. These test series indicated that gradients of $E_{acc}$ ~ 35 MV/m can be reliably reached in ICHIRO-type cavities.

The high gradient program at KEK was designed to demonstrate the capability to achieve ILC gradients of $E_{acc}$ = 35 MV/m or higher gradients in 9-cell cavities, initially fabricated from fine grain material. Therefore three 9-cell ICHIRO-type cavities were fabricated (I9#9, I9#10, and I9 #11) and I9#9 was tested several times. The fabrication process for I9#10 and I9#11 incorporated an improved electron beam welding procedure by welding the equators from the inside of the cavity cell. This provides better surface finishes and avoids irregularities in the weld, which often appear at full penetration welds from the outside. Presently, only I9#9 has been tested several times and is limited at $E_{acc}$ = 27 MV/m after horizontal BCP, this treatment method was developed to achieve better uniformity of material removal and therefore better field flatness in the 9-cell cavity. More tests on the 9-cell cavities are planned for the future. Also, a collaborative effort with Tokyo Denkai is aimed at producing single crystal ingots by using a seed crystal on which the ingot will grow [69].

## 4.4 Michigan State University (MSU), USA

At Michigan State University in collaboration with Fermi Lab a strong material science program was pursued, focusing on gaining a better understanding of the influence of material properties of niobium and cavity preparation methods on cavity performance. The advent of large grain niobium was welcomed as an additional interesting research topic [93]. The ongoing collaboration between MSU and FNAL also included the design and manufacturing of prototype cavities for the proposed Proton Driver, originally planned to be made from fine grain material. Large grain niobium opened the possibility of comparing both material and in collaboration with JLab; MSU fabricated two single cell cavities from CBMM niobium provided by JLab. These two cavities were also treated and tested at JLab, whereas two simultaneously fabricated cavities from fine grain niobium were evaluated at MSU. The encouraging results of these efforts are displayed in Fig. 32:

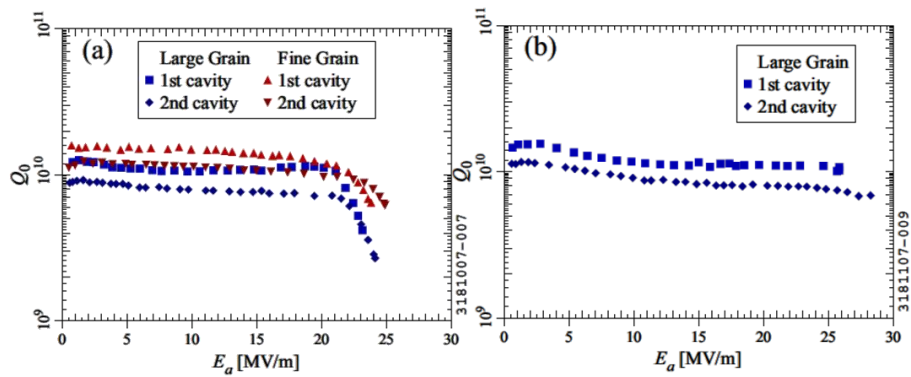

Figure 32: Test results for single cell cavities: (a) fine grain cavities received BCP and HPR treatment, large grain cavities were post-purified and then BCP and high pressure rinsed and (b) large grain cavity after additional bake out [94].

Besides the single cell program, two 7-cell cavities were fabricated one of each from fine grain and large grain material (Fig. 33). The cavities were completed in 2007 but no test results are available to the authors' knowledge.

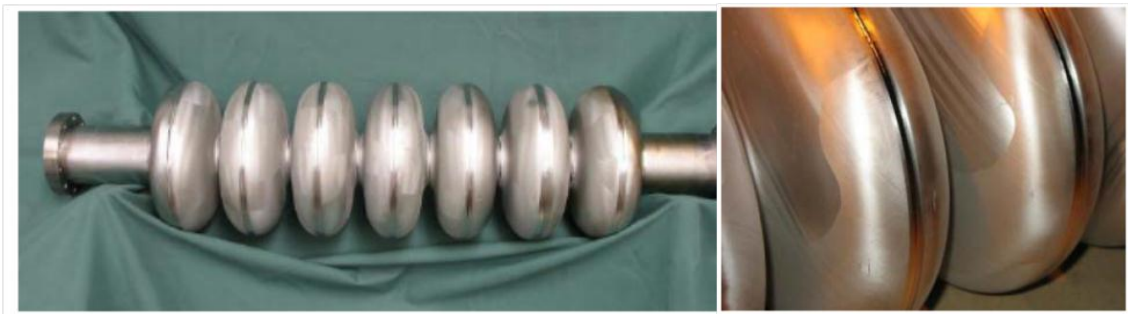

Figure 33: Prototype 7-cell 1300 MHz cavity for the proton driver made from large grain niobium [94].

## 4.5 Cornell University, USA

The interest at Cornell University in large grain niobium centered on the question of the causes for non-linearities in the $Q_0$ vs $E_{acc}$ behavior of a cavity. Therefore, comparative studies were made on cavities made from both fine grain and large grain niobium using temperature maps. Large grain material with few, but distinct grain boundaries allows to look at correlations between grain boundary patterns and heating pattern on the outside of the cavity at higher rf fields [95]. Such a map superimposed on the grain boundary pattern is shown on Fig. 34.

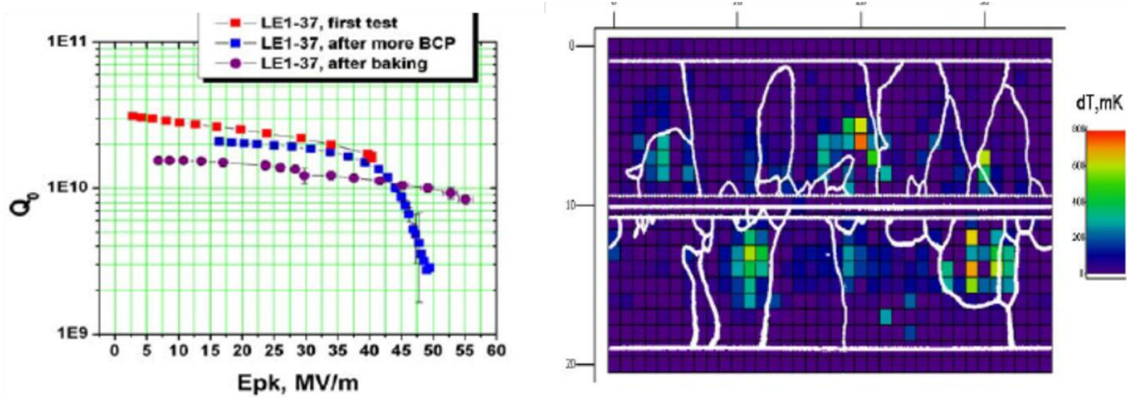

Figure 34: Temperature map of a large grain 1500 MHz single cell cavity superimposed on the grain boundary pattern (white lines). The map was taken at a peak electric field of $E_p = 49$ MV/m [95].

Figure 34 clearly indicates that there is no correlation between grain boundaries and heated areas: even the breakdown spot is not on a grain boundary, but was identified by optical inspection as a geometrical defect such as a pit. This is an important result reinforcing the notion that grain boundary steps caused by preferential etching is not necessarily causing quench limitations. The result of a comparison to a similar behaving small grain cavity with a $Q$-slope as pointed out in [96] is that the heating in the $Q$-slope area is spread further out in fine grain material than in larger grain niobium.

Another important contribution to understanding the non-linearities in the $Q_0$ vs $E_{acc}$ behavior of niobium cavities has been reported in [95, 96]. The approach taken in this work was as following:
- Two cavities from fine grain and large grain material were tested after BCP surface treatment and temperature maps were taken.
- With the T-maps areas of "hot spots" and "cold spots" were identified in the Q-drop region.
- These cavity areas were cut out and analyzed with surface analytical methods such as EBSD (electron back-scattered diffraction), XPS (X-ray photoelectron spectroscopy), AES (Auger electron spectroscopy) and optical profilometry.

The result of this investigation is summarized below;

- Surface roughness is not the cause for the non-linearities.
- Grain boundaries could be eliminated.
- Oxide structure in the niobium/oxide interface did not show any difference between "cold" and "hot" spots.
- However, EBSD maps, which are measuring mis-orientations between neighboring crystals and are an indirect measure of crystal defects such as vacancies and dislocations revealed large differences between hot spots and cold spots (see Fig. 35) and might be responsible for the observed non-linearities. As discussed in [45] it attracts residual hydrogen from the surface of the niobium and can form lossy compounds of NbH.

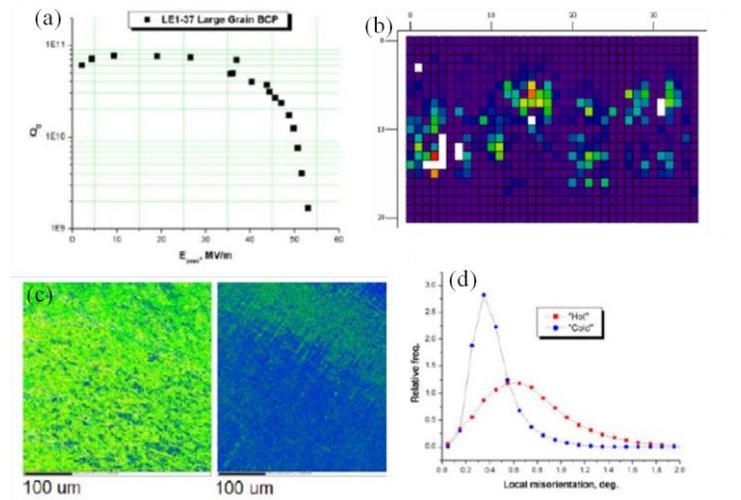

Figure 35: (a) Performance of large grain cavity, (b) T-map with "hot" and "cold" areas, (c) compares an EBDS map of a hot spot (green) and a cold spot (blue) and (d) the difference in occurrence of local mis-orientations between hot and cold areas [96].

### 4.6 Peking University, China

Peking University started the studies of the large grain cavities in 2005. One 1.3 GHz single cell and one 2-cell TESLA shaped cavity were fabricated with the large grain sheet from Ningxia. Both cavities were tested at JLab with excellent performances; the single cell cavity reached an accelerating gradient of $E_{acc}$>40 MV/m with $Q_0$ of ~ $10^{10}$ as shown in Fig. 36 [97]. Because of this outstanding result some surface inspections were done at FNAL making use of the replica technique developed at this lab [36]. Some of the astonishing features discovered at the cavity surface are shown in Fig. 37. That led to the conclusion that surface smoothness seems to be only a secondary effect determining the performance of a cavity.

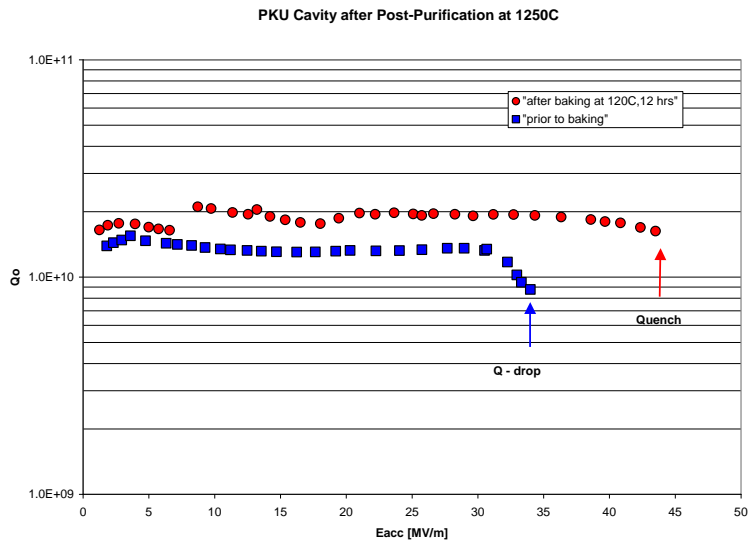

Figure 36: Large grain single cell cavity of the TESLA shape [from ref. 97].

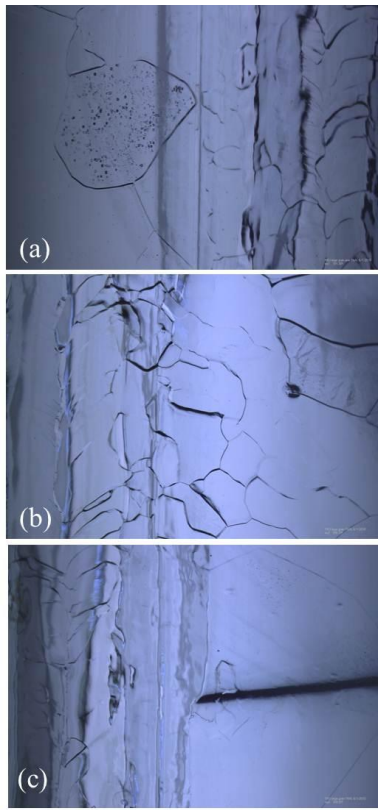

Figure 37: The optical image of the PKU-LG1 single cell showing (a) BCP etch pits, (b) pits in weld, and (c) the BCP enhanced grain boundary a measured enhancement of 1.6 [from ref.36].

At a later date, a 3.5-cell photo injector cavity were fabricated at PKU and tested at JLab. It reached a gradient of $E_{acc}$ = 23.5 MV/m, and a $Q$–value higher than $1.2 \times 10^{10}$ as shown in Fig. 38.

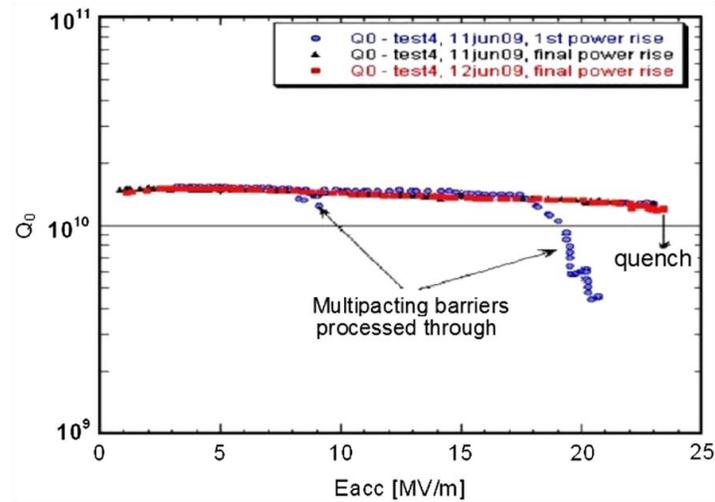

Figure 38: Final test result of 3.5-cell large grain cavity after 800 °C heat treatment and additional 30 μm BCP followed by HPR [98].

All cavities achieved respectable results with gradients $E_{acc}$ > 20 MV/m, however, several of the electron beam welds at equators of the cells showed rather large defects. It seems likely, that with improved electron beam welding the cavities will perform to higher levels. In 2008 PKU produced a single crystal single cell cavity using the same enlargement method as developed at DESY and described earlier. The cavity was evaluated at JLab. Initially the performance was rather poor due to a sub-standard electron beam weld at the equator. After several iterations of mechanical grinding of the weld and subsequent chemical surface treatment by BCP the performance improved significantly and a gradient of $E_{acc}$~ 29 MV/m was reached.

### 4.7 IHEP, China

IHEP started R&D on the 1.3 GHz large grain cavity in 2006. Three electro-polished ICHIRO single cell cavities were fabricated and processed in KEK with Ningxia large grain niobium material provided by IHEP. The maximum gradient achieved was 47.9 MV/m [99]. Later two LL type single cell cavities were fabricated at IHEP and tested at KEK which reached the accelerating gradient of ~40 MV/m [100]. The first 9-cell cavity IHEP-01 was fabricated and processed at IHEP and tested at the KEK. The maximum gradient reached ~20 MV/m, as shown in Fig.39 [101].

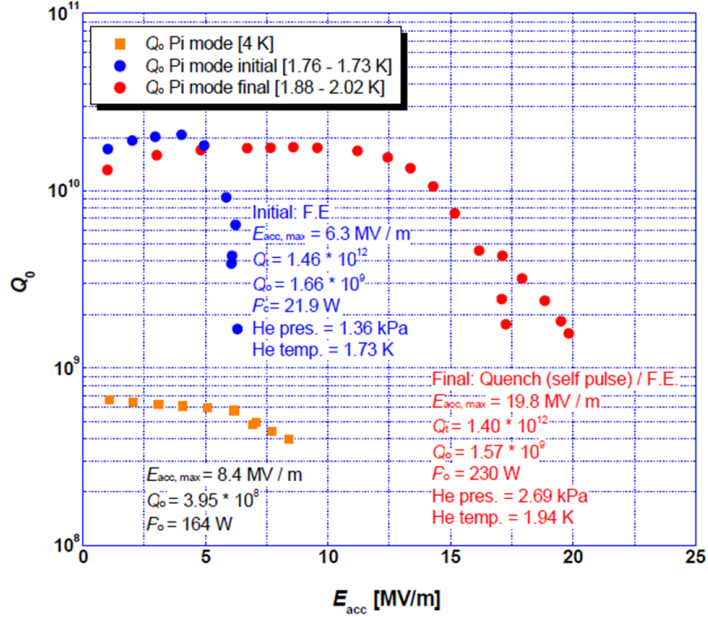

Figure 39: Test result of the IHEP-01 large grain los-loss shape 9-cell cavity (without HOM couplers) [101].

### 4.8 BARC, India

Bhabha Atomic Research Centre (BARC) recently started the activities on cavity fabrications targeting to develop the accelerator technology for the accelerator driven subcritical system. A single cell cavity with the resonance frequency of 1050 MHz, $\beta = 0.49$ was fabricated from the ingot provided by CBMM with RRR ~ 96. The cavity was tested at JLab and quenched at an accelerating gradient of ~ 6 MV/m ($B_p$ ~ 48 mT) with low field $Q_0$ of ~ $10^{10}$ [102]. This "sub-standard" performance is suspected to be caused by a large number of etch pits observed after the chemical etching.

## 5. DISCUSSION

There are at the present two applications for superconducting cavities in accelerators:
- Accelerators such as ERL's or proton accelerators are operated in CW where only moderate gradients ($E_{acc} \leq 20$ MV/m) and high $Q$–values are needed to reduce the cryogenic losses.
- Very high energy accelerators such as ILC need very high gradients ($E_{acc}$>30 MV/m), but since they are operating in a pulsed mode, high $Q$-values at these gradients are not as important as in the CW case.

For SRF technology these applications represent several challenges:

- Understand the causes of losses in a cavity/material and in particular understand and eliminate the reasons for "residual losses".
- Understand the causes for "premature" quenches in a cavity and eliminate the causes.
- Understand the non-linear losses especially at high surface fields and eliminate them or shift them to higher gradients.

For a long time grain boundaries have been suspected as the causes for both higher losses and premature field limitations, since grain boundaries represent a "weak" link with degraded superconducting properties because of preferential segregation of impurities such as oxygen, hydrogen, and carbon. In electromagnetic fields at rf frequencies currents are forced to flow across these boundaries, which will suffer a transition to the normal conducting state at field levels below the bulk critical fields. The magneto-optical imaging work and measurements of transport properties on large grain bi-crystals have confirmed this picture as discussed above. However, it was also found that magnetic flux penetration and flux flow is significant only if the external magnetic field is parallel to the grain boundaries. Since the grain boundary orientation in an rf cavity is rather random, this situation does also occur randomly as has been shown by the T-maps overlayed with the grain boundary. The heating and quench does not necessarily occur at a grain boundary at high fields. It is also not obvious whether the results regarding flux entry and flux flow obtained by dc techniques (magneto-optic and transport methods) are directly applicable to the case where an rf field is applied.

Cavities from ingot niobium have only ~1/100 of the length of grain boundaries than for cavities of identical shape made of fine-grain Nb. This might be one of the reasons why the residual resistance in large grain material is generally lower than in cavities made from fine grain niobium. It has been shown in earlier experiments at Saclay [103] that the resistance across a grain boundary is the major contribution to the local RRR-value and can be a factor of 1000 larger than commonly assumed. Similar measurements on a bi-crystal showed a degradation of the RRR−value across the grain boundary by 11 %, indicating again that there are grain boundary contributions to resistivity.

If the grain boundaries contribute to the residual resistance, than one should expect the lowest residual resistance in single crystal cavities. This however is not the case, even when these cavities have gone through the same treatments. Other mechanisms must contribute to the residual resistance. Therefore there is inconclusive evidence regarding the role of grain boundaries on both *Q*−value and gradient limitations.

RF niobium cavities, whether fabricated from fine grain, large grain or single crystal material, still reach only in rare cases magnetic surface field levels close to the theoretical limit. This inferior performance have been linked to localized defects in the material near electron beam welds by powerful surface inspection optical systems or by replica techniques as demonstrated in ref. [35, 36].The mechanism of how these defects, pits or protrusions develop is not totally clear. However, there is some evidence that large numbers of crystal defects such as dislocations, vacancies or residual strain fields are preferentially attacked by the acids and form etch pits. Such surface topographical features have been correlated with "hot spot" behavior in the cavities. Point contact tunneling experiments showed that the "hot" samples have degraded superconducting properties (lower energy gap and a larger pair breaking parameter) compared to "cold" samples There are, in principle, may possible causes for such degraded superconducting properties, as pointed out in [104], ranging from niobium hydrides to localized magnetic moments within the surface oxide, to sub-oxide precipitates.

. The material studies on the large grain/single crystal material using magneto-optical magnetization, susceptibility, penetration depth or laser scanning microscopy methods have explored the influence of surface treatment on trapped magnetic flux. The results confirm the possibility of pinned vortices as a source of rf losses in Nb and that large-grain has weaker pinning than fine-grain. The detailed nature of the pinning in this "technical Nb" material is not yet clear.

Besides improvement in the quality factor and accelerating gradient of SRF cavities for future accelerators, the material cost is another important issue. Ingot Nb offers the possibility of reducing the material cost, particularly if higher tantalum content and lower RRR, compared to what is specified for fine-grain Nb, will be accepted.. The quench field for SRF cavities is connected with the high thermal conductivity via the RRR of the materials. However, the presence of the phonon peak in the thermal conductivity of ingot niobium at the operating temperature ~2K make the SRF cavities fabricated from ingot niobium superior, from the thermal stability point of view, to the high RRR fine grain niobium. Results on BCP treated 9-cell ingot Nb cavities showed no impact on the quench field for RRR values ranging from 150 to 500 [105]. Also, the tests of single-cell cavities showed no correlation between Ta content, between 150-1300 wt. ppm, and maximum $E_{acc}$ up to ~30 MV/m [72].

## 6. SUMMARY

Ingot niobium was re-introduced in 2004 as the result of collaboration between Jefferson Lab and CBMM, one of the largest niobium producing companies in the world. The "excitement" for this material was generated by the promise of producing similar performance levels in SRF cavities as with fine grain niobium at a lower price. This expectation was based on a potentially "streamlined" material production process, a less elaborate cavity manufacturing process and a reduced and less expensive surface preparation technique. Most of these expectations have been fulfilled as described above, even though not all of them. For example, a greater demand for ingot Nb discs is needed from cavity manufacturers in order to decrease the cost of slicing per disc. An accelerator such as the ILC, which would need more than $10^5$ niobium sheets, could provide such opportunity. Also, presently ingot niobium is only available at a diameter of app. 18" due to the size of existing electron beam melting facilities. This diameter limits the use of this material for SRF cavities up to a frequency of ~750 MHz. However, larger melting facilities are being considered by the industry.

Another issue, which remains to be fully settled, is the mechanical stability of the material at room temperature and cryogenic temperatures related to the pressure vessel code. The "Deutsches Elektronen Synchrotron" (DESY) has taken the lead in settling this issue by incorporating two large grain cavities in cryomodules operating in the FLASH accelerator and plans to use a full cryomodule of 9-cell cavities made from ingot niobium into its X-FEL accelerator. The cavities in FLASH are operating without problems. DESY in collaboration with W. C. Heraeus has developed specifications for ingot niobium and a production process which can create niobium sheet material with a large single crystal in the center of the disc and a specified orientation. Such material was successfully used at DESY to fabricate twelve 9-cell X-FEL type cavities with industry. However, the DESY specifications [67] in terms of RRR and Ta content are the same as those for fine-grain Nb. As have results in other laboratories shown, excellent performances can also be achieved with somewhat relaxed specifications: no large center crystal, higher Ta contents and lower RRR value. As mentioned above the specified mechanical properties are in compliance with the pressure vessel code especially if the overpressure during cool-down is reduced to < 4 bar Additional niobium producers will be able to meet less stringent specifications. Besides these two outstanding issues the introduction of ingot niobium into the SRF community has generated an enormous amount of interest especially in the material scientist and surface science communities and the authors confidently can attest that ingot niobium is arguably one of the most thoroughly investigated technical materials.

It should not be forgotten that the highest accelerating gradient in a multi-cell cavity has been achieved with ingot niobium, corresponding to a surface magnetic field of $H_{peak}$~192 mT close to the theoretical critical field of niobium. Also, ingot niobium is less "lossy" than fine grain material, which translates into higher $Q$-values at medium accelerating gradients. In this respect, an exceptionally high $Q_0$ (~5×10$^{10}$ at 1.5 GHz, 2.0 K and 90 mT) with ingot Nb. Such results make ingot Nb cavities highly desirable for both high-gradient pulsed accelerators as well as for CW accelerators. It is our conclusion that ingot niobium has many advantages over poly-crystalline material and it is ready and matured enough to be used in the next generation of accelerating machines.

## ACKNOWLEDGEMENT

The Jefferson Lab author's would like to thank T. Carneiro of CBMM for his enthusiastic support of the development of this technology in the initial stages and also Dr. Swapan Chattopadhy, the former director of the accelerator division at JLab for his far sighted view of these activities and his important support.